\documentclass{mscs}

\usepackage{comment} % for auxproof
\newif\ifshowauxproof
\showauxprooffalse
\newcommand{\auxprooffont}{\small}
\makeatletter
\newcommand{\auxproof}[1]{%
\ifshowauxproof
{\auxprooffont
\textbf{AuxP} #1 \textbf{AuxQ}}%
\else\@bsphack\@esphack\fi}
\makeatother

\ifshowauxproof
\newenvironment{Auxproof}
  {\par\auxprooffont\noindent\textbf{AUX-PROOF}\dotfill\par
    \noindent\ignorespaces}
  {\par\noindent\textbf{AUX-QED}\dotfill\par
    \noindent\ignorespacesafterend}
\else
\excludecomment{Auxproof}
\fi

\usepackage[utf8]{inputenc}
\usepackage[T1]{fontenc}
\usepackage{textcomp}
\usepackage{lmodern}
\usepackage{exscale} % load this between lmodern and amsmath

\usepackage{graphicx,color}
\usepackage{hyperref}
\usepackage{etoolbox}
\usepackage{xparse}
\usepackage{xspace}

\usepackage{latexsym}
\usepackage{amssymb,amsmath}
\usepackage{mathtools}
\usepackage{stmaryrd}
\usepackage{bussproofs}
\usepackage{nicefrac}

\usepackage[noadjust]{cite}
\usepackage{enumitem}
\usepackage{microtype}
\usepackage{wrapfig}
\usepackage{listings} 
\usepackage{fancybox}

\usepackage{todonotes}
\presetkeys{todonotes}{size=\footnotesize,color=orange!30}{}
\allowdisplaybreaks[1] % amsmath option: allows page breaks inside a
\usepackage{xy}
\xyoption{all}
\xyoption{tips}
\SelectTips{cm}{}  % Tips (of arrows) are in accordance with Computer Modern
\usepackage{tikz}
\usetikzlibrary{circuits.ee.IEC}
\usetikzlibrary{shapes,shapes.geometric}

\newlength\stateheight
\newlength\minimumstatewidth
\tikzset{width/.initial=\minimummorphismwidth}
\tikzset{colour/.initial=white}

\newif\ifblack\pgfkeys{/tikz/black/.is if=black}
\newif\ifwedge\pgfkeys{/tikz/wedge/.is if=wedge}
\newif\ifvflip\pgfkeys{/tikz/vflip/.is if=vflip}
\newif\ifhflip\pgfkeys{/tikz/hflip/.is if=hflip}
\newif\ifhvflip\pgfkeys{/tikz/hvflip/.is if=hvflip}

\pgfdeclarelayer{foreground}
\pgfdeclarelayer{background}
\pgfsetlayers{background,main,foreground}

\def\thickness{0.4pt}

\makeatletter
\pgfkeys{%
  /tikz/on layer/.code={
    \pgfonlayer{#1}\begingroup
    \aftergroup\endpgfonlayer
    \aftergroup\endgroup
  },
  /tikz/node on layer/.code={
    \gdef\node@@on@layer{%
      \setbox\tikz@tempbox=\hbox\bgroup\pgfonlayer{#1}\unhbox\tikz@tempbox\endpgfonlayer\pgfsetlinewidth{\thickness}\egroup}
    \aftergroup\node@on@layer
  },
  /tikz/end node on layer/.code={
    \endpgfonlayer\endgroup\endgroup
  }
}
\def\node@on@layer{\aftergroup\node@@on@layer}
\makeatother

\makeatletter
\pgfdeclareshape{state}
{
    \savedanchor\centerpoint
    {
        \pgf@x=0pt
        \pgf@y=0pt
    }
    \anchor{center}{\centerpoint}
    \anchorborder{\centerpoint}
    \saveddimen\overallwidth
    {
        \pgf@x=3\wd\pgfnodeparttextbox
        \ifdim\pgf@x<\minimumstatewidth
            \pgf@x=\minimumstatewidth
        \fi
    }
    \savedanchor{\upperrightcorner}
    {
        \pgf@x=.5\wd\pgfnodeparttextbox
        \pgf@y=.5\ht\pgfnodeparttextbox
        \advance\pgf@y by -.5\dp\pgfnodeparttextbox
    }
    \anchor{A}
    {
        \pgf@x=-\overallwidth
        \divide\pgf@x by 4
        \pgf@y=0pt
    }
    \anchor{B}
    {
        \pgf@x=\overallwidth
        \divide\pgf@x by 4
        \pgf@y=0pt
    }
    \anchor{text}
    {
        \upperrightcorner
        \pgf@x=-\pgf@x
        \ifhflip
            \pgf@y=-\pgf@y
            \advance\pgf@y by 0.4\stateheight
        \else
            \pgf@y=-\pgf@y
            \advance\pgf@y by -0.4\stateheight
        \fi
    }
    \backgroundpath
    {
       \begin{pgfonlayer}{foreground}
        \pgfsetstrokecolor{black}
        \pgfsetlinewidth{\thickness}
        \pgfpathmoveto{\pgfpoint{-0.5*\overallwidth}{0}}
        \pgfpathlineto{\pgfpoint{0.5*\overallwidth}{0}}
        \ifhflip
            \pgfpathlineto{\pgfpoint{0}{\stateheight}}
        \else
            \pgfpathlineto{\pgfpoint{0}{-\stateheight}}
        \fi
        \pgfpathclose
        \ifblack
            \pgfsetfillcolor{black!50}
            \pgfusepath{fill,stroke}
        \else
            \pgfusepath{stroke}
        \fi
       \end{pgfonlayer}
    }
}
\makeatother

\tikzset{inline text/.style =
  {text height=1.2ex, text depth=0.25ex,yshift=0.5mm}}
\tikzset{arrow box/.style =
  {rectangle,inline text,fill=white,draw,
    minimum height=5mm,yshift=-0.5mm,minimum width=5mm}}
\tikzset{dot/.style =
  {inner sep=0mm,minimum width=1mm,minimum height=1mm,
    draw,shape=circle}}
\tikzset{white dot/.style = {dot,fill=white,text depth=-0.2mm}}
\tikzset{scalar/.style = {diamond,draw,inner sep=1pt}}

\tikzset{copier/.style = {dot,fill,text depth=-0.2mm}}
\tikzset{discarder/.style = {circuit ee IEC,ground,rotate=90,scale=0.8,xshift=.55ex}}

\tikzset{xshiftu/.style = {shift = {(#1, 0)}}}
\tikzset{yshiftu/.style = {shift = {(0, #1)}}}

\newenvironment{myproof}[1][Proof ]%
   { \begin{trivlist}%
     \item[\hskip \labelsep {\bfseries #1}]%
   }%
   { \end{trivlist}%
   }

\makeatother
 \newdir{ >}{{}*!/-7.5pt/@{>}}
 \newdir{|>}{!/4.5pt/@{|}*:(1,-.2)@^{>}*:(1,+.2)@_{>}}
 \newdir{ |>}{{}*!/-3pt/@{|}*!/-7.5pt/:(1,-.2)@^{>}*!/-7.5pt/:(1,+.2)@_{>}}
\newcommand{\xyline}[2][]{\ensuremath{\smash{\xymatrix@1#1{#2}}}}
\newcommand{\xyinline}[2][]{\ensuremath{\smash{\xymatrix@1#1{#2}}}^{\rule[8.5pt]{0pt}{0pt}}}
\newcommand{\filter}{\raisebox{5.5pt}{$\xymatrix@=6pt@H=0pt@M=0pt@W=4pt{\\ \ar@{>->}[u]}$}}
\newcommand{\ideal}{\raisebox{1pt}{$\xymatrix@=5pt@H=0pt@M=0pt@W=4pt{\ar@{>->}[d] \\ \mbox{}}$}}
\makeatletter

\newcommand{\QEDbox}{\square}
\newcommand{\QED}{\hspace*{\fill}$\QEDbox$}

\newcommand{\after}{\mathrel{\circ}}

\newcommand{\idmap}[1][]{\ensuremath{\mathrm{id}_{#1}}}

\newcommand{\EfProb}{\textrm{EfProb}\xspace}

\newcommand{\Pow}{\mathcal{P}}

\newcommand{\Dst}{\mathcal{D}}
\newcommand{\Giry}{\mathcal{G}}

\newcommand{\UF}{\ensuremath{\mathcal{U}{\kern-.75ex}\mathcal{F}}}

\newcommand{\NNO}{\mathbb{N}}

\newcommand{\R}{\mathbb{R}}

\newcommand{\closed}{\ensuremath{\mathcal{C}{\kern-.45ex}\ell}}
\newcommand{\ket}[1]{\ensuremath{|{\kern.1em}#1{\kern.1em}\rangle}}
\newcommand{\bigket}[1]{\ensuremath{\big|#1\big\rangle}}

\newcommand{\Kl}{\mathcal{K}{\kern-.5ex}\ell}
\newcommand{\set}[2]{\{#1\;|\;#2\}}

\newcommand{\andthen}{\ensuremath{\mathrel{\&}}}

\newcommand{\tuple}[1]{\langle#1\rangle}

\newcommand{\intd}{{\kern.2em}\mathrm{d}{\kern.03em}}
\newcommand{\indic}[1]{\mathbf{1}_{#1}}

\newcommand{\dst}{\ensuremath{\mathsf{dst}}}
\newcommand{\st}{\ensuremath{\mathsf{st}}}
\newcommand{\leftScottint}{[{\kern-.3ex}[}
\newcommand{\rightScottint}{]{\kern-.3ex}]}

\newcommand{\betachan}{\ensuremath{\mathrm{Beta}}}
\newcommand{\flipchan}{\ensuremath{\mathrm{Flip}}}
\newcommand{\normchan}{\ensuremath{\mathrm{Norm}}}
\newcommand{\binomchan}{\ensuremath{\mathrm{Binom}}}
\newcommand{\dirchan}{\ensuremath{\mathrm{Dir}}}
\newcommand{\multchan}{\ensuremath{\mathrm{Mult}}}
\newcommand{\copychan}{\ensuremath{\mathrm{copy}}}

\newsavebox\sbground
\savebox\sbground{%
  \begin{tikzpicture}[baseline=0pt,circuit ee IEC]
    \draw (0,-.1ex) to (0,.85ex)
    node[ground,point up,scale=.6,xshift=.55ex] {};
  \end{tikzpicture}%
}

\newsavebox\sbcopier
\savebox\sbcopier{%
\begin{tikzpicture}[baseline=0pt]
\node[copier,scale=.7] (a) at (0,3.6pt) {};
\draw (a) -- +(-90:.16);
\draw (a) -- +(45:.19);
\draw (a) -- +(135:.19);
\end{tikzpicture}}
\newcommand{\copier}{\mathord{\usebox\sbcopier}}

\DeclareFixedFont{\ttb}{T1}{txtt}{bx}{n}{12} % for bold
\DeclareFixedFont{\ttm}{T1}{txtt}{m}{n}{12}  % for normal
\usepackage{color}
\definecolor{deepblue}{rgb}{0,0,0.5}
\definecolor{deepred}{rgb}{0.6,0,0}
\definecolor{deepgreen}{rgb}{0,0.5,0}

\newcommand\pythonstyle{\lstset{
backgroundcolor = \color{lightgray},
language=Python,
basicstyle=\ttm,
otherkeywords={self,>>>},             % Add keywords here
keywordstyle=\ttb\small\color{deepblue},
emph={@,+\%%
literate={.+}{{{\color{red}.+}}}2 {.**}{{{\color{red}.\**{}}}}2 {*}{{{\color{red}*}}}1
},          % Custom highlighting
emphstyle=\ttb\small\color{deepred},    % Custom highlighting style
stringstyle=\small\color{deepgreen},
frame=tb,                         % Any extra options here
showstringspaces=false            % 
}}
\lstnewenvironment{python}[1][] % Python environment
{
\pythonstyle
\lstset{basicstyle=\normalfont\ttfamily\small#1}
}
{}

\newcommand\pythoninline[1]{\pythonstyle\lstinline[basicstyle=\normalfont\ttfamily\small]{#1}} % Python for inline

\newtheorem{theorem}{Theorem}[section]
\newtheorem{proposition}[theorem]{Proposition}
\newtheorem{lemma}[theorem]{Lemma}
\newtheorem{corollary}[theorem]{Corollary}

\newtheorem{definition}[theorem]{Definition}

\newtheorem{example}[theorem]{Example}

\renewcommand{\arraystretch}{1.3}
\title[A Channel-Based Perspective on Conjugate Priors]{A
  Channel-Based Perspective on Conjugate Priors\thanks{The research
    leading to these results has received funding from the European
    Research Council under the European Union's Seventh Framework
    Programme (FP7/2007-2013) / ERC grant agreement nr.~320571}}

\author[B. Jacobs]{%
  B\ls A\ls R\ls T\ns
  J\ls A\ls C\ls O\ls B\ls S\\
  Institute for Computing and Information Sciences
  Radboud University\addressbreak
  P.O.Box 9010, 6500 GL Nijmegen, the Netherlands}

\begin{document}

\maketitle

\begin{abstract} 
A desired closure property in Bayesian probability is that an updated
posterior distribution be in the same class of distributions --- say
Gaussians --- as the prior distribution. When the updating takes place
via a statistical model, one calls the class of prior distributions
the `conjugate priors' of the model.  This paper gives (1)~an abstract
formulation of this notion of conjugate prior, using channels, in a
graphical language, (2)~a simple abstract proof that such conjugate
priors yield Bayesian inversions, and (3)~a logical description of
conjugate priors that highlights the required closure of the priors
under updating. The theory is illustrated with several standard
examples, also covering multiple updating.
\end{abstract}

\section{Introduction}\label{sec:intro}

The main result of this paper, Theorem~\ref{thm:conjugateinversion},
is mathematically trivial. But it is not entirely trivial to see that
this result is trivial. The effort and contribution of this paper lies
in setting up a framework --- using the abstract language of channels,
Kleisli maps, and string diagrams for probability theory --- to define
the notion of conjugate prior in such a way that there is a trivial
proof of the main statement, saying that conjugate priors yield
Bayesian inversions. This is indeed what conjugate priors are meant to
be.

%% http://lesswrong.com/lw/5sn/the_joys_of_conjugate_priors/

%% Suppose you haven't observed any coin flips yet, but you have some
%% intuition about what the distribution should be.  Then you can choose
%% values for and that represent your prior understanding of the coin.
%% Higher values of indicate more confidence in your intuition; thus,
%% choosing the appropriate hyperparameters is a method of quantifying
%% your prior understanding so that it can be used in computation.  and
%% will act like "imaginary data"; when you update your distribution over
%% after observing a coin flip , it will be like you already saw heads
%% and tails before that coin flip.

%% In conclusion, the beta distribution, which is a conjugate prior to
%% the bernoulli and binomial distributions, is super awesome.  It makes
%% it possible to do Bayesian reasoning in a computationally efficient
%% manner, as well as having the philosophically satisfying
%% interpretation of representing real or imaginary prior data.

% Overview of conjugate prior families:
%
% https://fisher.osu.edu/~schroeder.9/AMIS900/ech6.pdf
%
% by: https://fisher.osu.edu/people/schroeder.9

Conjugate priors form a fundamental topic in Bayesian theory.  They
are commonly described via a closure property of a class of prior
distributions, namely as being closed under certain Bayesian updates.
Conjugate priors are especially useful because they do not only
involve a closure \emph{property}, but also a particular
\emph{structure}, namely an explicit function that performs an
analytical computation of posterior distributions via updates of the
parameters. This greatly simplify Bayesian analysis. For instance, the
$\betachan$ distribution is conjugate prior to the Bernoulli (or
`flip') distribution, and also to the binomial distribution: updating
a $\betachan(\alpha,\beta)$ prior via a Bernoulli/binomial statistical
model yields a new $\betachan(\alpha',\beta')$ prior, with adapted
parameters $\alpha',\beta'$ that can be computed explicitly from
$\alpha,\beta$ and the observation at hand. Despite this importance,
the descriptions in the literature of what it means to be a conjugate
prior are remarkably informal.  One does find several lists of classes
of distributions, for instance at Wikipedia\footnote{See
  \url{https://en.wikipedia.org/wiki/Conjugate_prior} or online lists,
  such as
  \url{https://www.johndcook.com/CompendiumOfConjugatePriors.pdf},
  consulted at Sept.\ 10, 2018}, together with formulas about how to
re-compute parameters.  The topic has a long and rich history in
statistics (see \textit{e.g.}~\cite{Bishop06}), with much emphasis on
exponential families~\cite{DiaconisY79}, but a precise, general
definition is hard to find.

We briefly review some common approaches, without any pretension to be
complete: the definition in~\cite{Alpaydin10} is rather short, based
on an example, and just says: ``We see that the posterior has the same
form as the prior and we call such a prior a \emph{conjugate prior}.''
Also~\cite{RussellN03} mentions the term `conjugate prior' only in
relation to an example. There is a separate section in~\cite{Bishop06}
about conjugate priors, but no precise definition. Instead, there is
the informal description ``\ldots the posterior distribution has the
same functional form as the prior.'' The most precise definition
(known to the author) is in~\cite[\S5.2]{BernardoS00}, where the
conjugate family with respect to a statistical model, assuming a
`sufficient statistic', is described. It comes close to our
channel-based description, since it explicitly mentions the conjugate
family as a conditional probability distribution with (re-computed)
parameters.  The approach is rather concrete however, and the high
level of mathematical abstraction that we seek here is missing
in~\cite{BernardoS00}.

%% We shall see that an important role is played by conjugate priors,
%% that lead to posterior distributions having the same functional form
%% as the prior, and that there- fore lead to a greatly simplified
%% Bayesian analysis.

This paper presents a novel systematic perspective for precisely
defining what conjugate priorship means, both via diagrams and via
(probabilistic) logic. It uses the notion of `channel' as starting
point. The basis of this approach lies in category theory, especially
effectus theory~\cite{Jacobs15d,ChoJWW15b}. However, we try to make
this paper accessible to non category theorists, by using the term
`channel' instead of morphism in a Kleisli category of a suitable
monad. Moreover, a graphical language is used for channels that
hopefully makes the approach more intuitive. Thus, the emphasis of the
paper is on \emph{what it means} to have conjugate priors.  It does
not offer new perspectives on how to find/obtain them.

The paper is organised as follows. It starts in
Section~\ref{sec:ideas} with a high-level description of the main
ideas, without going into technical details. Preparatory definitions
are provided in Sections~\ref{sec:Kleisli} and~\ref{sec:inversion},
dealing with channels in probabilistic computation, with a
diagrammatic language for channels, and with Bayesian inversion.
Then, Section~\ref{sec:conjugate} contains the novel channel-based
definition of conjugate priorship; it also illustrates how several
standard examples fit in this new
setting. Section~\ref{sec:conjugatepriorinversion} establishes the
(expected) close relationship between conjugate priors and Bayesian
inversions. Section~\ref{sec:logic} then takes a fresh perspective, by
re-describing the Bayesian-inversion based formulation in more logical
terms, using validity and updating. This re-formulation captures the
intended closure of a class of priors under updating in the most
direct manner. It is used in Section~\ref{sec:multiple} to illustrate
how multiple updates are handled, typically via a `sufficient
statistic'.

\section{Main ideas}\label{sec:ideas}

This section gives an informal description of the main ideas
underlying this paper. It starts with a standard example, and then
proceeds with a step-by-step introduction to the essentials of the
perspective of this paper.

\begin{figure}
\begin{center}
\includegraphics[width=0.3\textwidth]{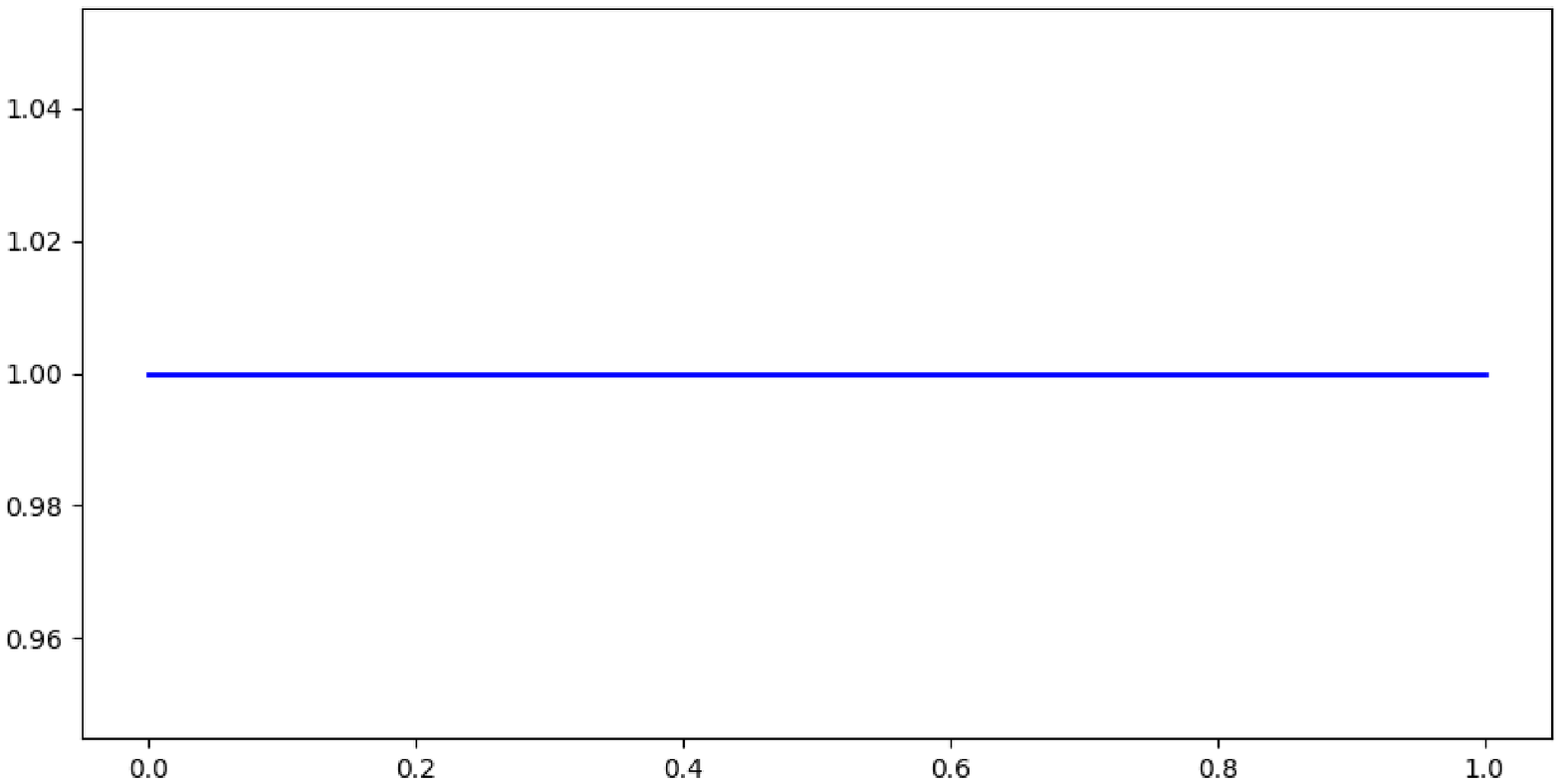}
\includegraphics[width=0.3\textwidth]{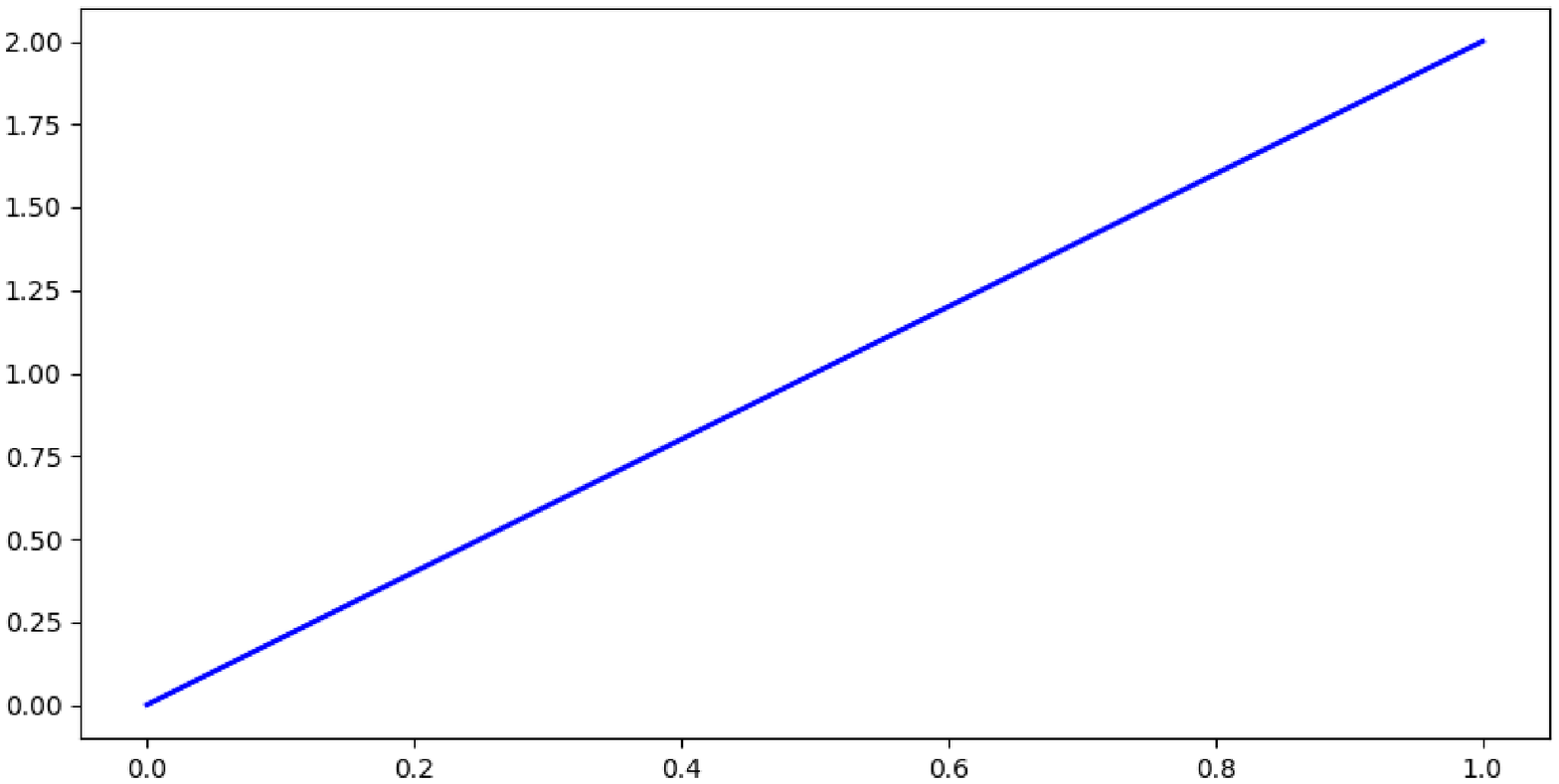}
\includegraphics[width=0.3\textwidth]{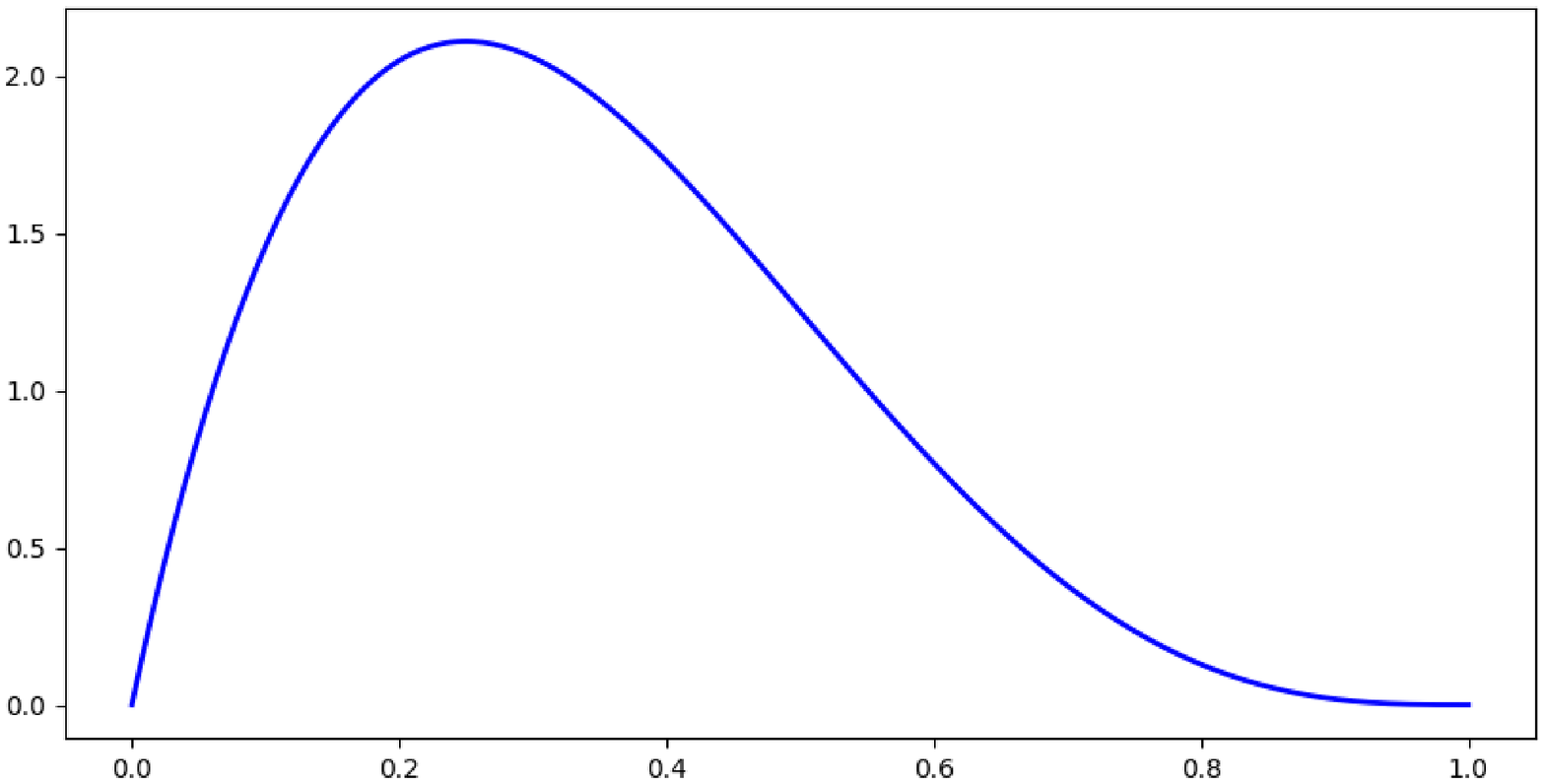}
\end{center}
\caption{Uniform prior, and two posterior probability density
  functions on $[0,1]$, after observing head, and after observing
  head-tail-tail-tail. These functions correspond respectively to
  $\betachan(1,1)$, $\betachan(2,1)$,
  $\betachan(2,4)$. Example~\ref{ex:efprobcoin} below explains how
  these three plots are obtained, via actual Bayesian updates
  (inversions), and not by simply using the $\betachan$ functions.}
\label{fig:coin}
\end{figure}

A well-known example in Bayesian reasoning is inferring the (unknown)
bias of a coin from a sequence of consecutive head/tail observations.
The bias is a number $r \in [0,1]$ in the unit interval, giving the
`Bernoulli' or `flip' probability $r$ for head, and $1-r$ for
tail. Initially we assume a uniform distribution for $r$, as described
by the constant probability density function (pdf) on the left in
Figure~\ref{fig:coin}.  After observing one head, this pdf changes to
the second picture. After observing head-tail-tail-tail we get the
third pdf. These pictures are obtained by Bayesian inversion, see
Section~\ref{sec:inversion}.

It is a well-known fact that all the resulting distributions are
instances of the $\betachan(\alpha,\beta)$ family of distributions,
for different parameters $\alpha,\beta$. After each observation, one
can re-compute the entire updated distribution, via Bayesian
inversion, as in Example~\ref{ex:efprobcoin}. But in fact there is a
much more efficient way to obtain the revised distribution, namely by
computing the new parameter values: increment $\alpha$ by one, for
head, and increment $\beta$ by one for tail, see
Examples~\ref{ex:betaflip} and~\ref{ex:betaflipupdate} for
details. The family of distributions $\betachan(\alpha,\beta)$,
indexed by parameters $\alpha,\beta$, is thus suitably closed under
updates with Bernoulli. It is the essence of the statement that
$\betachan$ is conjugate prior to Bernoulli. This will be made precise
later on.

Let $X = (X, \Sigma)$ be a measurable space, where $\Sigma \subseteq
\Pow(X)$ is a $\sigma$-algebra of measurable subsets. We shall write
$\Giry(X)$ for the set of probability distributions on $X$. Elements
$\omega\in\Giry(X)$ are thus countably additive functions
$\omega\colon\Sigma \rightarrow [0,1]$ with $\omega(X) = 1$.
\begin{description}
\item[Idea 1:] A \emph{family} of distributions on $X$, indexed by a
  measurable space $P$ of parameters, is a (measurable) function $P
  \rightarrow \Giry(X)$.  Categorically, such a function is a
  \emph{Kleisli} map for $\Giry$, considered as monad on the category
  of measurable spaces (see Section~\ref{sec:Kleisli}). These Kleisli
  maps are also called \emph{channels}, and will be written simply as
  arrows $P \rightarrow X$, or diagrammatically as boxes
  \ $\scriptstyle\vcenter{\hbox{%
\begin{tikzpicture}[font=\tiny]
\node[arrow box, scale=0.5] (c) at (0,0) {$\;$};
\draw (c) to (0,0.35);
\draw (c) to (0,-0.35);
\node at (0.2,0.3) {$X$};
\node at (0.2,-0.3) {$P$};
\end{tikzpicture}}}$ where we imagine that information is flowing upwards.
\end{description}

\noindent The study of families of distributions goes back a long way,
\textit{e.g.} as `experiments'~\cite{Blackwell51}.

Along these lines we shall describe the family of $\betachan$
distributions as a channel with $P = \R_{>0}\times\R_{>0}$ and $X =
[0,1]$, namely as function:
\begin{equation}
\label{diag:beta}
\vcenter{\xymatrix@C+1pc{
\R_{>0}\times\R_{>0}\ar[r]^-{\betachan} & \Giry([0,1])
}}
\end{equation}

\noindent For $(\alpha,\beta) \in \R_{>0}\times\R_{>0}$ there is the
probability distribution $\betachan(\alpha,\beta) \in \Giry([0,1])$
determined by its value on a measurable subset $M\subseteq [0,1]$,
which is obtained via integration:
\begin{equation}
\label{eqn:beta}
\begin{array}{rcl}
\betachan(\alpha,\beta)(M)
& = &
{\displaystyle\int_{M}}
   \frac{x^{\alpha-1}(1-x)^{\beta-1}}{B(\alpha,\beta)}\intd x,
\end{array}
\end{equation}

\noindent where $B(\alpha,\beta) = \int_{[0,1]}
x^{\alpha-1}(1-x)^{\beta-1} \intd x$ is a normalisation constant.

A conjugate prior relationship involves a family of distributions $P
\rightarrow \Giry(X)$ which is closed wrt.\ updates based on
observations (or: data) from a separate domain $O$. Each `parameter'
element $x\in X$ gives rise to a separate distribution on $O$. This is
what is usually called a \emph{statistical} or \emph{parametric}
model. We shall also describe it as a channel.

%% Such a likelihood on spaces $X,O$ is a parameterised probability
%% density function (pdf), consisting of a function $\ell \colon X\times
%% O\rightarrow \R_{\geq 0}$ satisfying $\int \ell(x,y)\intd y = 1$ for
%% each $x\in X$.

\begin{description}
\item[Idea 2:] The observations for a family $P \rightarrow \Giry(X)$
  arise via another ``Kleisli'' map $X \rightarrow \Giry(O)$
  representing the statistical model. Conjugate priorship will be
  defined for two such composable channels $P \rightarrow X
  \rightarrow O$, where $O$ is the space of observations.
\end{description}

%% \noindent The channel $X\rightarrow \Giry(O)$ associated with a
%% likelihood $\ell \colon X\times O \rightarrow \R_{\geq 0}$ is given on
%% an element $x\in X$ and a measurable subset $M\subseteq O$ as
%% $\int_{M}\ell(x,y)\intd y$.

In the above coin example, the space $O$ of observations is the
two-element set $2 = \{0,1\}$ where $0$ is for tail and $1$ for head.
The Bernoulli channel is written as $\flipchan \colon [0,1]
\rightarrow \Giry(2)$.  A probability $r\in [0,1]$ determines a
Bernoulli/flip/coin probability distribution $\flipchan(r) \in
\Giry(2)$ on $2$, formally sending the subset $\{1\}$ to $r$ and
$\{0\}$ to $1-r$.

\begin{description}
\item[Idea 3:] A channel $c\colon P\rightarrow X$ is a conjugate prior
  to a channel $d\colon X \rightarrow O$ if there is a \emph{parameter
    translation function} $h\colon P\times O \rightarrow P$ satisfying
  a suitable equation.
\end{description}

\noindent The idea is that $c(p)$ is a prior, for $p\in P$, which gets
updated via the statistical model (channel) $d$, in the light of
observation $y\in O$. The revised, updated distribution is
$c(h(p,y))$. The model $d$ is usually written as a conditional
probability $d(y\mid \theta)$.

In the coin example we have $h\colon \R_{>0}\times\R_{>0} \times 2
\rightarrow \R_{>0}\times\R_{>0}$ given by $h(\alpha,\beta,1) =
(\alpha+1,\beta)$ and $h(\alpha,\beta,0) = (\alpha,\beta+1)$, see
Example~\ref{ex:betaflip} below for more information.

What has been left unexplained is the `suitable' equation that the
parameter translation function $h\colon P\times O \rightarrow P$
should satisfy. It is not entirely trivial, because it is an equation
between channels in what is called the Kleisli category $\Kl(\Giry)$
of the Giry monad $\Giry$. At this stage we need to move to a more
categorical description.  The equation, which will appear in
Definition~\ref{def:conjugateprior}, bears similarities with the
notion of Bayesian inversion, which will be introduced in
Section~\ref{sec:inversion}.

\section{Channels and conditional probabilities}\label{sec:Kleisli}

This section will describe conditional probabilities as arrows and
will show how to compose them. Thereby we are entering the world of
category theory. We aim to suppress the underlying categorical
machinery and make this work accessible to readers without such
background. For those with categorical background knowledge: we will
be working in the Kleisli categories of the distribution monad $\Dst$
for discrete probability, and of the Giry monad $\Giry$ for continuous
probability, see \textit{e.g.}~\cite{Giry82,Panangaden09,Jacobs17a}.
Discrete distributions may be seen as a special case of continuous
distributions, via a suitable inclusion map $\Dst \rightarrow \Giry$.
Hence one could give one account, using $\Giry$ only. However, in
computer science, unlike for instance in statistics, discrete
distributions are so often used that they merit separate treatment.

We thus start with discrete probability. We write a (finite, discrete)
distribution on a set $X$ as a formal convex sum $r_{1}\ket{x_1} +
\cdots + r_{n}\ket{x_n}$ of elements $x_{i}\in X$ and probabilities
$r_{i}\in [0,1]$ with $\sum_{i}r_{i}=1$. The `ket' notation $\ket{-}$
is syntactic sugar, used to distinguish elements of $x$ from their
occurrence $\ket{x}$ in such formal convex sums\footnote{Sometimes
  these distributions $\sum_{i}r_{i}\ket{x_i}$ are called
  `multinomial' or `categorical'; the latter terminology is confusing
  in the present context.}. A distribution as above can be identified
with a `probability mass' function $\omega \colon X \rightarrow [0,1]$
which is $r_{i}$ on $x_{i}$ and $0$ elsewhere. We often implicitly
identify distributions with such functions. We shall write $\Dst(X)$
for the set of distributions on $X$.

We shall focus on functions of the form $c\colon X \rightarrow
\Dst(Y)$.  They give, for each element $x\in X$ a distribution $c(x)$
on $Y$. Hence such functions form an $X$-indexed collection
$\big(c(x)\big)_{x\in X}$ of distributions $c(x)$ on $Y$. They can be
understood as \emph{conditional} probabilities $P(y\mid x) = r$, if
$c(x)$ is of the form $\cdots r\ket{y}\cdots$, with weight $r =
c(x)(y)\in[0,1]$ for $y\in Y$.  Thus, by construction, $\sum_{y}
P(y\mid x) = 1$, for each $x\in X$.  Moreover, if the sets $X$ and $Y$
are finite, we can describe $c\colon X \rightarrow \Dst(Y)$ as a
stochastic matrix, with entries $P(y\mid x)$, adding up to one --- per
row or column, depending on the chosen orientation of the matrix.

We shall often write functions $X \rightarrow \Dst(Y)$ simply as
arrows $X \rightarrow Y$, call them `channels', and write them as
`boxes' in diagrams. This arrow notation is justified, because there
is a natural way to compose channels, as we shall see shortly.  But
first we describe \emph{state transformation}, also called
\emph{prediction}. Given a channel $c\colon X \rightarrow \Dst(Y)$ and
a state $\omega\in\Dst(X)$, we can form a new state, written as $c \gg
\omega$, on $Y$. It is defined as:
\begin{equation}
\label{eqn:discstatransf}
\begin{array}{rcl}
c \gg \omega
& \coloneqq &
{\displaystyle\sum_y} \big(\sum_{x} \omega(x)\cdot c(x)(y)\big)\bigket{y}.
\end{array}
\end{equation}

\noindent The outer sum $\sum_{y}$ is a formal convex sum, whereas the
inner sum $\sum_{x}$ is an actual sum in the unit interval $[0,1]$.
Using state transformation $\gg$ it is easy to define composition of
channels: given functions $c\colon X \rightarrow \Dst(Y)$ and $d\colon
Y \rightarrow \Dst(Z)$, we use the ordinary composition symbol
$\after$ to form a composite channel $d \after c \colon X \rightarrow
\Dst(Z)$, where:
\begin{equation}
\label{eqn:discretecomposition}
\begin{array}{rcccl}
(d \after c)(x)
& \coloneqq &
d \gg c(x)
& = &
{\displaystyle\sum_{z\in Z}} \big(\sum_{y} c(x)(y)\cdot d(y)(z)\big)
   \bigket{z}.
\end{array}
\end{equation}

\noindent Essentially, this is matrix composition for stochastic
matrices. Channel composition $\after$ is associative, and also has a
neutral element, namely the identity channel $\eta \colon X
\rightarrow X$ given by the `Dirac' function $\eta(x) = 1\ket{x}$.  It
is not hard to see that $(d \after c) \gg \omega = d \gg (c \gg
\omega)$.

\medskip

We turn to channels in continuous probability. As already mentioned in
Section~\ref{sec:ideas}, we write $\Giry(X)$ for the set of
probability distributions $\omega\colon \Sigma_{X} \rightarrow [0,1]$,
where $X = (X,\Sigma_{X})$ is a measurable space. These probability
distributions are (also) called states.  The set $\Giry(X)$ carries a
$\sigma$-algebra itself, but that does not play an important role
here. Each element $x\in X$ yields a probability measure
$\eta(x)\in\Giry(X)$, with $\eta(x)(M) = \indic{M}(x)$, which is $1$
if $x\in M$ and $0$ otherwise. This map $\indic{M} \colon X
\rightarrow [0,1]$ is called the indicator function for the subset
$M\in\Sigma_{X}$.

For a state/measure $\omega\in\Giry(X)$ and a measurable function
$f\colon X \rightarrow \R_{\geq 0}$ we write $\int f\intd \omega$ for
the Lebesgue integral, if it exists. We follow the notation
of~\cite{Jacobs13a} and refer there for details, or alternatively,
to~\cite{Panangaden09}. We recall that an integral $\int_{M} f\intd
\omega$ over a measurable subset $M\subseteq X$ of the domain of $f$
is defined as $\int \indic{M}\cdot f \intd\omega$, and that $\int
\indic{M} \intd\omega = \omega(M)$. Moreover, $\int f \intd\eta(x) =
f(x)$.

For a measurable function $g\colon X\rightarrow Y$ between measurable
spaces $X,Y$ there is the `push forward' function $\Giry(g) \colon
\Giry(X) \rightarrow \Giry(Y)$, given by $\Giry(g)(\omega)(N) =
\omega\big(g^{-1}(N)\big)$. It satisfies:
\begin{equation}
\label{eqn:invmeasure}
\begin{array}{rcl}
\displaystyle\int f \intd \Giry(g)(\omega)
& = &
\int f \after g \intd \omega.
\end{array}
\end{equation}

\noindent Often, the measurable space $X$ is a subset $X\subseteq\R$
of the real numbers and a probability distribution $\omega$ on $X$ is
given by a probability density function (pdf), that is, by a
measurable function $f\colon X \rightarrow \R_{\geq 0}$ with $\int_{X}
f(x) \intd x = 1$. Such a pdf $f$ gives rise to a state $\omega
\in\Giry(X)$, namely:
\begin{equation}
\label{eqn:statefrompdf}
\begin{array}{rcl}
\omega(M)
& = &
\displaystyle \int_{M} f(x) \intd x.
\end{array}
\end{equation}

\noindent We then write $\omega = \int f$.

In this continuous context a channel is a measurable function $c\colon
X \rightarrow \Giry(Y)$, for measurable spaces $X,Y$. Like in the
discrete case, it gives an $X$-indexed collection
$(\big(c(x)\big)_{x\in X}$ of probability distributions on $Y$. 
The channel $c$ can transform a state $\omega\in\Giry(X)$ on $X$ 
into a state $c \gg \omega \in\Giry(Y)$ on $Y$, given on a measurable
subset $N\subseteq Y$ as:
\begin{equation}
\label{eqn:statetransformation}
\begin{array}{rcl}
\big(c \gg \omega\big)(N)
& = &
\displaystyle\int c(-)(N) \intd \omega.
\end{array}
\end{equation}

\noindent For another channel $d\colon Y \rightarrow \Giry(Z)$ there
is a composite channel $d \after c \colon X \rightarrow \Giry(Z)$, via
integration:
\begin{equation}
\label{eqn:continuouscomposition}
\begin{array}{rcccl}
\big(d \after c\big)(x)(K)
& \coloneqq &
\big(d \gg c(x)\big)(K)
& = &
\displaystyle\int d(-)(K) \intd c(x)
\end{array}
\end{equation}

In many situations a channel $c\colon X \rightarrow \Giry(Y)$ is given
by an indexed probability density function (pdf) $u\colon X \times Y
\rightarrow \R_{\geq 0}$, with $\int u(x,y)\intd y = 1$ for each $x\in
X$.  The associated channel $c$ is:
\begin{equation}
\label{eqn:channelfrompdf}
\begin{array}{rcl}
c(x)(N)
& = &
\displaystyle \int_{N} u(x,y) \intd y.
\end{array}
\end{equation} 

\noindent In that case we simply write $c = \int u$ and call $c$ a
pdf-channel. We have already seen such a description of the
$\betachan$ distribution as a pdf-channel in~\eqref{eqn:beta}.

(In these pdf-channels $X \rightarrow Y$ we use a collection of pdf's
$u(x,-)$ which are all dominated by the Lebesgue measure. This
domination happens via the relationship $\ll$ of absolute continuity,
using the Radon-Nikodym Theorem, see
\textit{e.g.}~\cite{Panangaden09}.)

Various additional computation rules for integrals are given in the
Appendix.

\section{Bayesian inversion in string diagrams}\label{sec:inversion}

In this paper we make superficial use of string diagrams to
graphically represent sequential and parallel composition of channels,
mainly in order to provide an intuitive visual overview. We refer
to~\cite{Selinger11} for mathematical details, and mention here only
the essentials.

A channel $X \rightarrow Y$, for instance of the sort discussed in
the previous section, can be written as a box \ $\scriptstyle\vcenter{\hbox{%
\begin{tikzpicture}[font=\tiny]
\node[arrow box, scale=0.5] (c) at (0,0) {$\;$};
\draw (c) to (0,0.35);
\draw (c) to (0,-0.35);
\node at (0.2,0.3) {$Y$};
\node at (0.2,-0.3) {$X$};
\end{tikzpicture}}}$ with information flowing upwards, from the
wire labeled with $X$ to the wire labeled with $Y$. Composition of
channels, as in~\eqref{eqn:discretecomposition}
or~\eqref{eqn:continuouscomposition}, simply involves connecting
wires (of the same type). The identity channel is just a wire.  
We use a triangle notation \ \raisebox{.3em}{$\scriptstyle\vcenter{\hbox{%
\begin{tikzpicture}[font=\tiny]
\node[state, scale=0.5] (omega) at (0,0) {$\;$};
\draw (omega) to (0,0.2);
\node at (0.15,0.15) {$X$};
\end{tikzpicture}}}$} for a state on $X$. It is special case of
a channel, namely of the form $1 \rightarrow X$ with trivial singleton
domain $1$.

In the present (probabilistic) setting we allow copying of wires,
written diagrammatically as $\copier$. We briefly describe such copy
channels for discrete and continuous probability:
\[ \xymatrix@R-1.8pc{
X\ar[r]^-{\copier} & \Dst(X\times X)
& &
X\ar[r]^-{\copier} & \Giry(X\times X)
\\
x\ar@{|->}[r] & 1\ket{x,x}
& &
x\ar@{|->}[r] & \big(M\times N \mapsto \indic{M\cap N}(x)\big)
} \]
After such a copy we can use parallel channels. We briefly describe
how this works, first in the discrete case. For channels $c\colon X
\rightarrow \Dst(Y)$ and $d\colon A \rightarrow \Dst(B)$ we have a
channel $c\otimes d \colon X\times A \rightarrow \Dst(Y\times B)$
given by:
\[ \begin{array}{rcl}
(c\otimes d)(x,a)
& = &
\displaystyle \sum_{y,b} c(x)(y)\cdot d(a)(b)\bigket{y,b}.
\end{array} \]
Similarly, in the continuous case, for channels $c\colon X
\rightarrow \Giry(Y)$ and $d\colon A \rightarrow \Giry(B)$ we
get $c\otimes d \colon X\times A \rightarrow \Giry(Y\times B)$ given
by:
\[ \begin{array}{rcl}
(c\otimes d)(x,a)(M\times N)
& = &
c(x)(M)\cdot d(a)(N).
\end{array} \]

\noindent Recall that the product measure on $Y\times B$ is generated
by measurable rectangles of the form $M\times N$, for $M\in\Sigma_{Y}$
and $N\in\Sigma_{B}$.

We shall use a tuple $\tuple{c,d}$ as convenient abbreviation for
$(c\otimes d) \after \copier$. Diagrammatically, parallel channels are
written as adjacent boxes.

We can now formulate what Bayesian inversion is. The definition is
couched in purely diagrammatic language, but is applied only to
probabilistic interpretations in this paper.

\begin{definition}
\label{def:inversion}
The \emph{Bayesian inversion} of a channel $c\colon X \rightarrow Y$
with respect to a state $\omega$ of type $X$, if it exists, is a
channel in the opposite direction, written as $c^{\dag}_{\omega}
\colon Y \rightarrow X$, such that the following equation holds.
\begin{equation}
\label{diag:inversion}
\vcenter{\hbox{%
\begin{tikzpicture}[font=\small]
\node[state] (omega) at (0,0) {$\omega$};
\node[copier] (copier) at (0,0.3) {};
\node[arrow box] (c) at (0.5,0.95) {$c$};
\coordinate (X) at (-0.5,1.5);
\coordinate (Y) at (0.5,1.5);
\draw (omega) to (copier);
\draw (copier) to[out=150,in=-90] (X);
\draw (copier) to[out=15,in=-90] (c);
\draw (c) to (Y);
\end{tikzpicture}}}
\qquad=\qquad
\vcenter{\hbox{%
\begin{tikzpicture}[font=\small]
\node[state] (omega) at (0,-0.55) {$\omega$};
\node[copier] (copier) at (0,0.3) {};
\node[arrow box] (d) at (-0.5,0.95) {$c^{\dag}_{\omega}$};
\coordinate (X) at (-0.5,1.5);
\coordinate (Y) at (0.5,1.5);
\node[arrow box] (c) at (0,-0.15) {$c$};
\draw (omega) to (c);
\draw (c) to (copier);
\draw (copier) to[out=165,in=-90] (d);
\draw (copier) to[out=30,in=-90] (Y);
\draw (d) to (X);
\end{tikzpicture}}}
\end{equation}
\end{definition}

The dagger notation $c^{\dag}_{\omega}$ is copied
from~\cite{ClercDDG17}. There the state $\omega$ is left implicit, via
a restriction to a certain comma category of kernels. In that setting
the operation $(-)^{\dag}$ is functorial, and forms a dagger category
(see \textit{e.g.}~\cite{AbramskyC09,Selinger07} for definitions).  In
particular, it preserves composition and identities of channels.
Equation~\eqref{diag:inversion} can also be written as:
$\tuple{\idmap, c} \gg \omega = \tuple{c^{\dag}_{\omega}, \idmap} \gg
(c \gg \omega)$. Alternatively, in the discrete case, with variables
explicit, it says: $c(x)(y)\cdot \omega(x) = c^{\dag}_{\omega}(y)(x)
\cdot (c \gg \omega)(y)$. This comes close to the `adjointness'
formulations that are typical for daggers.

Bayesian inversion gives a channel-based description of Bayesian
(belief) updates. We briefly illustrate this for the coin example from
Section~\ref{sec:ideas}, using the $\EfProb$ language~\cite{ChoJ17b}.

\begin{example}
\label{ex:efprobcoin}
In Section~\ref{sec:ideas} we have seen the channel $\flipchan \colon
[0,1] \rightarrow 2$ that sends a probability $r\in [0,1]$ to the coin
state $\flipchan(r) = r\ket{1} + (1-r)\ket{0}$ with bias $r$. The
Bayesian inversion $2 \rightarrow [0,1]$ of this channel performs a
belief update, after a head/tail observation. Without going into
details we briefly illustrate how this works in the $\EfProb$ language
via the following code fragment. The first line describes a channel
\pythoninline{Flip} of type $[0,1] \rightarrow 2$, where $[0,1]$ is
represented as \pythoninline{R(0,1)} and $2 = \{0,1\}$ as
\pythoninline{bool\_dom}. The expression \pythoninline{flip(r)} 
captures a coin with bias \pythoninline{r}.
\begin{python}
>>> Flip = chan_fromklmap(lambda r: flip(r), R(0,1), bool_dom)
>>> prior = uniform_state(R(0,1))
>>> w1 = Flip.inversion(prior)(True)
>>> w2 = Flip.inversion(w1)(False)
>>> w3 = Flip.inversion(w2)(False)
>>> w4 = Flip.inversion(w3)(False)
\end{python}

\noindent The (continuous) states \pythoninline{w1} --
\pythoninline{w4} are obtained as successive updates of the uniform
state \pythoninline{prior}, after successive observations
\pythoninline{True}-\pythoninline{False}-\pythoninline{False}-\pythoninline{False},
for head-tail-tail-tail. The three probability density functions in
Figure~\ref{fig:coin} are obtained by plotting the prior state, and
also the two states \pythoninline{w1} and \pythoninline{w4}.
\end{example}

It is relatively easy to define Bayesian inversion in discrete
probability theory: for a channel $c\colon X \rightarrow \Dst(Y)$ and
a state/distribution $\omega\in\Dst(X)$ one can define a channel
$c^{\dag}_{\omega} \colon Y \rightarrow \Dst(X)$ as:
\begin{equation}
\label{eqn:discreteinversion}
\begin{array}{rcccl}
c^{\dag}_{\omega}(y)(x)
& = &
\displaystyle\frac{\omega(x)\cdot c(x)(y)}{(c \gg \omega)(y)}
& = &
\displaystyle\frac{\omega(x)\cdot c(x)(y)}{\sum_{z}\omega(z)\cdot c(z)(y)},
\end{array}
\end{equation}

\noindent assuming that the denominator is non-zero. This corresponds
to the familiar formula $P(B\mid A) = \nicefrac{P(A,B)}{P(A)}$ for
conditional probability. The state $c^{\dag}_{\omega}(y)$ can
alternatively be defined via updating the state $\omega$ with the
point predicate $\{y\}$, transformed via $c$ into a predicate $c \ll
\indic{\{y\}}$ on $X$, see Section~\ref{sec:logic}
(and~\cite{JacobsZ16}) for details.

\begin{Auxproof}
Just to be sure:
\[ \begin{array}{rcl}
\big(\tuple{c^{\dag}_{\omega}, \idmap} \gg (c \gg \omega)\big)(x,y)
& = &
c^{\dag}_{\omega}(y)(x)\cdot (c \gg \omega)(y)
\\
& = &
\displaystyle\frac{\omega(x)\cdot c(x)(y)}{c_{*}(\omega)(y)} \cdot
   (c \gg \omega)(y)
\\
& = &
\omega(x)\cdot c(x)(y)
\\
& = &
(\tuple{\idmap,c} \gg \omega)(x,y).
\end{array} \]

\noindent And:
\[ \begin{array}{rcccl}
\omega|_{c \ll \indic{\{y\}}}(x)
& = &
\displaystyle\frac{\omega(x) \cdot (c \ll \indic{\{y\}})(x)}{\omega 
   \models c \ll \indix{\{y\}}}
& = &
\displaystyle\frac{\omega(x)\cdot c(x)(y)}{\sum_{x}\omega(x)\cdot c(x)(y)},
\end{array} \]
\end{Auxproof}

The situation is much more difficult in continuous probability theory,
since Bayesian inversions may not exist~\cite{AckermanFR11,Stoyanov14}
or may be determined only up to measure zero. But when restricted to
\textit{e.g.}~standard Borel spaces, as in~\cite{ClercDDG17},
existence is ensured, see also~\cite{Faden85,CulbertsonS14}. Another
common solution is to assume that we have a pdf-channel: there is a
map $u\colon X\times Y \rightarrow \R_{\geq 0}$ that defines a channel
$c \colon X \rightarrow \Giry(Y)$, like in~\eqref{eqn:channelfrompdf},
as $c(x)(N) = \int_{N} u(x,y)\intd y$.  Then, for a distribution
$\omega\in\Giry(X)$ we can take as Bayesian inversion:
\begin{equation}
\label{eqn:continuousinversion}
\begin{array}{rcl}
c^{\dag}_{\omega}(y)(M)
& = &
\displaystyle\frac{\int_{M} u(-,y)\intd \omega}{\int_{X} u(-,y)\intd \omega}
\\[+1em]
& = &
\displaystyle\frac{\int_{M} f(x)\cdot u(x,y) \intd x}
   {\int_{X} f(x) \cdot u(x,y) \intd x}
   \qquad \mbox{when } \omega = \int f(x)\intd x.
\end{array}
\end{equation}

\noindent We prove that this definition satisfies the 
inversion Equation~\eqref{diag:inversion}, using the calculation
rules from the Appendix.
\[ \begin{array}{rcl}
\big(\tuple{c^{\dag}_{\omega}, \idmap} \gg (c \gg \omega)\big)(M\times N)
& \smash{\stackrel{\eqref{eqn:statetransformation}}{=}} &
\displaystyle \int \tuple{c^{\dag}_{\omega}, \idmap}(-)(M\times N) 
   \intd (c \gg \omega)
\\
& \smash{\stackrel{(\ref{eqn:pdfintegration},
     \ref{eqn:pdfstatetransformation})}{=}} &
\displaystyle\int \big(\int f(x) \cdot u(x,y) \intd x\big) \cdot 
   \tuple{c^{\dag}_{\omega}, \idmap}(y)(M\times N) \intd y
\\
& \smash{\stackrel{\eqref{eqn:graphequation}}{=}} &
\displaystyle\int \big(\int f(x) \cdot u(x,y) \intd x\big) \cdot 
   c^{\dag}_{\omega}(y)(M) \cdot \indic{N}(y) \intd y
\\
& \smash{\stackrel{\eqref{eqn:continuousinversion}}{=}} &
\displaystyle\int_{N} \big(\int f(x) \cdot u(x,y) \intd x\big) \cdot 
   \frac{\int_{M} f(x)\cdot u(x,y) \intd x}
   {\int f(x)\cdot u(x,y) \intd x} \intd y
\\[+0.8em]
& = &
\displaystyle\int_{N} \int_{M} f(x) \cdot u(x,y) \intd x \intd y
\\
& \smash{\stackrel{\eqref{eqn:pdfgraphstatetransformation}}{=}} &
(\tuple{\idmap, c} \gg \omega)(M\times N).
\end{array} \]

\section{Conjugate priors}\label{sec:conjugate}

We now come to the core of this paper. As described in the
introduction, the informal definition says that a class of
distributions is conjugate prior to a statistical model if the
associated posteriors are \emph{in the same class} of distributions.
The posteriors can be computed via Bayesian
inversion~\eqref{eqn:continuousinversion} of the statistical model.

This definition of `conjugate prior' is a bit vague, since it loosely
talks about `classes of distributions', without further
specification. As described in `Idea 1' in Section~\ref{sec:ideas}, we
interpret `class of states on $X$' as channel $P\rightarrow X$, where
$P$ is the type of parameters of the class.

We have already seen this channel-based description for the class
$\betachan$ distributions, in~\eqref{diag:beta}, as channel $\betachan
\colon \R_{>0} \times \R_{>0} \rightarrow [0,1]$. This works more
generally, for instance for Gaussian (normal) distributions
$\normchan(\mu, \sigma)$, where $\mu$ is the mean parameter and
$\sigma$ is the standard deviation parameter, giving a channel of the
form:
\begin{equation}
\label{diag:normal}
\vcenter{\xymatrix@C+1pc{
\R\times\R_{>0}\ar[r]^-{\normchan} & \Giry(\R)
}}
\end{equation}

\noindent It is determined by its value on a measurable subset
$M\subseteq \R$ as the standard integral:
\begin{equation}
\label{eqn:normal}
\begin{array}{rcl}
\normchan(\mu,\sigma)(M)
& = &
\displaystyle\int_{M}
   \frac{1}{\sqrt{2\pi}\sigma}e^{-\frac{(x-\mu)^{2}}{2\sigma^{2}}}\intd x
\end{array}
\end{equation}

Given a channel $c\colon P \rightarrow X$, we shall look at states
$c(p)$, for parameters $p\in P$, as priors. The statistical model, for
which these $c(p)$'s will be described as conjugate priors, goes from
$X$ to some other object $O$ of `observations'. Thus our starting
point is a pair of (composable) channels the form:
\begin{equation}
\label{diag:conjugatechannels}
\vcenter{\xymatrix{
P\ar[r]^-{c} & X\ar[r]^-{d} & O
}}
\qquad\qquad\mbox{or, as diagram,}\qquad\qquad
\vcenter{\hbox{%
\begin{tikzpicture}[font=\small]
\node[arrow box] (c) at (0.0,0.5) {$c$};
\node[arrow box] (d) at (0.0,1.2) {$d$};
\coordinate (P) at (0.0,0.0);
\coordinate (Y) at (0.0,1.7);
\draw (P) to (c);
\draw (c) to (d);
\draw (d) to (Y);
\end{tikzpicture}}}
\end{equation}

\noindent Such a pair of composable channels may be seen as a 2-stage
hierarchical Bayesian model. In that context the parameters $P$ are
sometimes called `hyperparameters', see
\textit{e.g.}~\cite{BernardoS00}.  There, esp.\ in Defn~5.6 of
conjugate priorship one can also distinguish two channels, written as
$p(\theta\mid\tau)$ and $p(x\mid\theta)$, corresponding respectively
to our channels $c$ and $d$. The $\tau$ form the hyperparameters.

In this setting we come to our main definition that formulates the
notion of conjugate prior in an abstract manner, avoiding classes of
distributions. It contains the crucial equation that was missing in
the informal description in Section~\ref{sec:ideas}.

All our examples of (conjugate prior) channels are maps in the Kleisli
category of the Giry monad, but the formulation applies more
generally. In fact, abstraction purifies the situation and shows the
essentials. The definition below speaks of `deterministic' channels,
between brackets. This part will be explained later on, in the
beginning of Section~\ref{sec:conjugatepriorinversion}.  It can be
ignored for now.

\begin{definition}
\label{def:conjugateprior}
In the situation~\eqref{diag:conjugatechannels} we call channel $c$ a
\emph{conjugate prior} to channel $d$ if there is a (deterministic)
channel $h\colon P\times O \rightarrow P$ for which the following
equation holds:
\begin{equation}
\label{eqn:conjugateprior}
\vcenter{\hbox{%
\begin{tikzpicture}[font=\small]
\node[arrow box] (c) at (0.0,-0.2) {$c$};
\node[copier] (copier) at (0,0.3) {};
\node[arrow box] (d) at (0.5,0.95) {$d$};
\coordinate (X) at (-0.5,1.5);
\coordinate (Y) at (0.5,1.5);
\draw (c) to (copier);
\draw (c) to (0.0,-0.7);
\draw (copier) to[out=150,in=-90] (X);
\draw (copier) to[out=15,in=-90] (d);
\draw (d) to (Y);
\end{tikzpicture}}}
\quad 
= 
\quad
\vcenter{\hbox{%
\begin{tikzpicture}[font=\small]
\node[copier] (copier1) at (0,0.3) {};
\node[copier] (copier2) at (0.25,2.0) {};
\coordinate (X) at (-0.5,3.8);
\coordinate (Y) at (0.5,3.8);
\node[arrow box] (c1) at (0.25,0.8) {$c$};
\node[arrow box] (d) at (0.25,1.5) {$d$};
\node[arrow box] (h) at (-0.5,2.6) {$\;\;h\;\;$};
\node[arrow box] (c2) at (-0.5,3.3) {$c$};
\draw (copier1) to (0.0,0.0);
\draw (copier1) to[out=150,in=-90] (h.240);
\draw (copier1) to[out=30,in=-90] (c1);
\draw (c1) to (d);
\draw (d) to (copier2);
\draw (copier2) to[out=165,in=-90] (h.305);
\draw (h) to (c2);
\draw (c2) to (X);
\draw (copier2) to[out=30,in=-90] (Y);
\end{tikzpicture}}}
\end{equation}

\noindent Equivalently, in equational form:
\[ \begin{array}{rcl}
\tuple{\idmap, d} \after c
& = &
((c\after h)\otimes\idmap) \after \tuple{\idmap, \copier \after d \after c}.
\end{array} \]
\end{definition}

The idea is that the map $h\colon P\times O \rightarrow P$ translates
parameters, with an observation from $O$ as additional
argument. Informally, one gets a posterior state $c(h(p,y))$ from the
prior state $c(p)$, given the observation $y\in O$. The power of this
`analytic' approach is that it involves simple re-computation of
parameters, instead of more complicated updating of entire
states. This will be illustrated in several standard examples below.

%% This $h$ is deterministic, that is, it is an actual function $P\times
%% O\rightarrow P$, and not a Kleisli map/channel $P\times O\rightarrow
%% \Giry(P)$.

The above Equation~\eqref{eqn:conjugateprior} is formulated in an
abstract manner --- which is its main strength. We will derive an
alternative formulation of Equation~\eqref{eqn:conjugateprior} for
pdf-channels. It greatly simplifies the calculations in examples.

\begin{lemma}
\label{lem:conjugatepriorpdf}
Consider composable channels $\smash{P \stackrel{c}{\rightarrow} X
  \stackrel{d}{\rightarrow} O}$, as in~\eqref{diag:conjugatechannels},
for the Giry monad $\Giry$, where $c\colon P\rightarrow \Giry(X)$ and
$d\colon X \rightarrow \Giry(O)$ are given by pdf's $u\colon P\times X
\rightarrow \R_{\geq 0}$ and $v \colon X\times O\rightarrow \R_{\geq
  0}$, as pdf-channels $c = \int u$ and $d = \int v$. Let $c$ be
conjugate prior to $d$, via a measurable function $h\colon P\times O
\rightarrow P$.

Equation~\eqref{eqn:conjugateprior} then amounts to, for an element
$p\in P$ and for measurable subsets $M\subseteq X$ and $N\subseteq O$,
\begin{equation}
\label{eqn:conjugatepriorint}
\begin{array}{rcl}
\lefteqn{\displaystyle\int_{N} \int_{M} u(p,x) \cdot v(x,y) \intd x\intd y}
\\[+0.8em]
& = &
\displaystyle \int_{N} \Big(\int u(p,x)\cdot v(x,y)\intd x\Big) \cdot 
   \Big(\int_{M} u(h(p,y),x) \intd x\Big) \intd y.
\end{array}
\end{equation}

\noindent In order to prove this equation, it suffices to prove that
the two functions under the outer integral $\int_{N}$ are equal, that
is, it suffices to prove for each $y\in O$,
\begin{equation}
\label{eqn:conjugatepriorfun}
\begin{array}{rcl}
\displaystyle\int_{M} u(p,x) \cdot v(x,y) \intd x
& = &
\displaystyle \Big(\int u(p,x)\cdot v(x,y)\intd x\Big) \cdot 
   \Big(\int_{M} u(h(p,y),x) \intd x\Big).
\end{array}
\end{equation}
This formulation will be used in the examples below.
\end{lemma}

\begin{myproof}
We extensively use the equations for integration from
Section~\ref{sec:Kleisli} and from the Appendix, in order to
prove~\eqref{eqn:conjugatepriorint}.  The left-hand-side of
Equation~\eqref{eqn:conjugateprior} gives the left-hand-side
of~\eqref{eqn:conjugatepriorint}:
\[ \begin{array}{rcccl}
\big(\tuple{\idmap, d} \after c\big)(p)(M\times N)
& \smash{\stackrel{\eqref{eqn:continuouscomposition}}{=}} &
\big(\tuple{\idmap, d} \gg c(p)\big)(M\times N)
& \smash{\stackrel{\eqref{eqn:pdfgraphstatetransformation}}{=}} &
\displaystyle \int_{N} \int_{M} u(p,x) \cdot v(x,y) \intd x \intd y.
\end{array} \]

\noindent Unravelling the right-hand-side of~\eqref{eqn:conjugateprior}
is a bit more work:
\[ \begin{array}{rcl}
\lefteqn{\big((c \after h)\otimes\idmap) \after
   \tuple{\idmap, \copier \after d \after c}\big)(p)(M\times N)}
\\
& \smash{\stackrel{\eqref{eqn:continuouscomposition}}{=}} &
\displaystyle \int (c \after h)\otimes\idmap)(-)(M\times N)
   \intd \tuple{\idmap, \copier \after d \after c}(p)
\\
& \smash{\stackrel{\eqref{eqn:graphequation}}{=}} &
\displaystyle \int ((c \after h)\otimes\idmap)(-)(M\times N)
   \intd \big(\eta(p) \otimes 
    (\copier \after d \after c)(p)\big)
\\
& \smash{\stackrel{\eqref{eqn:productpdfintegration}}{=}} &
\displaystyle \int \int ((c \after h)\otimes\idmap)(-,-)(M\times N)
   \intd \eta(p) \intd (\copier \after d \after c)(p)\big)
\\
& = &
\displaystyle \int ((c \after h)\otimes\idmap)(p,-)(M\times N)
   \intd \Giry(\copier)(d \gg c(p))
\\
& \smash{\stackrel{\eqref{eqn:invmeasure}}{=}} &
\displaystyle \int ((c \after h)\otimes\idmap)(p,\copier(-))(M\times N)
   \intd (d \gg c(p))
\\
& \smash{\stackrel{(\ref{eqn:pdfintegration},
     \ref{eqn:pdfseqcomposition})}{=}} &
\displaystyle \int \Big(\int u(p,x)\cdot v(x,y)\intd x\Big) \cdot 
   ((c \after h)\otimes\idmap)(p,y,y)(M\times N) \intd y
\\
& = &
\displaystyle \int \Big(\int u(p,x)\cdot v(x,y)\intd x\Big) \cdot 
   c(h(p,y))(M) \cdot \indic{N}(y) \intd y
\\
& = &
\displaystyle \int_{N} \Big(\int u(p,x)\cdot v(x,y)\intd x\Big) \cdot 
   \Big(\int_{M} u(h(p,y),x) \intd x\Big) \intd y.
\end{array} \]

\noindent By combining this outcome with the earlier one we get the
desired equation~\eqref{eqn:conjugatepriorint}. \QED
\end{myproof}

One can reorganise Equation~\eqref{eqn:conjugatepriorfun} as
a normalisation fraction:
\begin{equation}
\label{eqn:conjugatepriorfunnorm}
\begin{array}{rcl}
\displaystyle\int_{M} u(h(p,y),x) \intd x
& = &
\displaystyle \frac{\int_{M} u(p,x) \cdot v(x,y) \intd x}
                   {\int u(p,x)\cdot v(x,y)\intd x}.
\end{array}
\end{equation}

\noindent It now strongly resembles
Equation~\eqref{eqn:continuousinversion} for Bayesian inversion. This
connection will be established more generally in
Theorem~\ref{thm:conjugateinversion}. Essentially, the above
normalisation fraction~\eqref{eqn:conjugatepriorfunnorm} occurs
in~\cite[Defn.~5.6]{BernardoS00}. Later, in Section~\ref{sec:logic} we
will see that~\eqref{eqn:conjugatepriorfunnorm} can also be analysed
in terms of updating a state with a random variable.

We are now ready to review some standard examples. The first one
describes the structure underlying the coin example in
Section~\ref{sec:ideas}.

\begin{example}
\label{ex:betaflip}
It is well-known that the beta distributions are conjugate prior to
the Bernoulli `flip' likelihood function. We shall re-formulate this
fact following the pattern of Definition~\ref{def:conjugateprior},
with two composable channels, as in~\eqref{diag:conjugatechannels},
namely:
\[\xymatrix@C+1pc{
\NNO_{>0}\times\NNO_{>0}\ar[r]^-{\betachan} & 
   [0,1]\ar[r]^-{\flipchan} & 2 
   \rlap{\qquad where $2=\{0,1\}$.}
} \qquad \]

\noindent The $\betachan$ channel is as in~\eqref{diag:beta}, but now
restricted to the non-negative natural numbers $\NNO_{>0}$. We recall
that the normalisation constant $B(\alpha,\beta)$ is $\int_{[0,1]}
x^{\alpha-1}(1-x)^{\beta-1} \intd x$.

The $\flipchan$ channel sends a probability $r\in[0,1]$ to the
$\text{Bernoulli}(r)$ distribution, which can also be written as a
discrete distribution $\flipchan(r) = r\ket{1} + (1-r)\ket{0}$.  More
formally, as a Kleisli map $[0,1]\rightarrow\Giry(2)$ it is, for a
subset $N\subseteq 2$,
\[ \begin{array}{rcccccl}
\flipchan(r)(N)
& = &
\displaystyle \int_{N} r^{i}\cdot (1-r)^{1-i} \intd i
& = &
\displaystyle\sum_{i\in N} r^{i}\cdot (1-r)^{1-i}
& = &
\left\{{\renewcommand{\arraystretch}{1.0}\begin{array}{ll}
0 & \mbox{if } N = \emptyset \\
r  & \mbox{if } N = \{1\} \\
1-r \;\; & \mbox{if } N = \{0\} \\
1 & \mbox{if } N = \{0,1\}.
\end{array}}\right.
\end{array} \]

\noindent The $i$ in $\intd i$ refers here to the counting measure.

In order to show that $\betachan$ is a conjugate prior of $\flipchan$
we have to produce a parameter translation function $h\colon
\NNO_{>0}\times\NNO_{>0}\times 2 \rightarrow
\NNO_{>0}\times\NNO_{>0}$. It is defined by distinguishing the
elements in $2 = \{0,1\}$
\begin{equation}
\label{eqn:betaflipfun}
\begin{array}{rclcrcl}
h(\alpha, \beta, 1)
& = &
(\alpha+1, \beta)
& \qquad\mbox{and}\qquad &
h(\alpha, \beta, 0)
& = &
(\alpha, \beta+1).
\end{array}
\end{equation}

\noindent Thus, in one formula, $h(\alpha,\beta,i) = (\alpha+i,
\beta+(1-i))$.

We prove Equation~\eqref{eqn:conjugatepriorfun} for $c = \betachan =
\int u$ and $d = \flipchan = \int v$. We start from its
right-hand-side, for an arbitrary $i\in 2$,
\[ \begin{array}{rcl}
\lefteqn{\displaystyle \Big(\int u(\alpha,\beta,x)\cdot v(x,i)\intd x\Big) 
   \cdot \Big(\int_{M} u(h(\alpha,\beta,i),x) \intd x\Big)}
\\
& = &
\displaystyle\Big(\int \frac{x^{\alpha-1}(1-x)^{\beta-1}}{B(\alpha,\beta)} \cdot 
      x^{i}\cdot (1-x)^{1-i} \intd x\Big) \cdot
   \Big(\int_{M} \frac{x^{\alpha+i-1}(1-x)^{\beta+(1-i)-1}}
      {B(\alpha+i,\beta+(1-i))} \intd x\Big)
\\[+1em]
& = &
\displaystyle\Big(\frac{\int x^{\alpha+i-1}(1-x)^{\beta+(1-i)-1} \intd x}
      {B(\alpha,\beta)}\Big) \cdot
   \Big(\int_{M} \frac{x^{\alpha-1}(1-x)^{\beta-1}}
      {B(\alpha+i,\beta+(1-i))} \cdot x^{i} \cdot (1-x)^{1-i} \intd x\Big)
\\[+1em]
& = &
\displaystyle\Big(\frac{B(\alpha+i,\beta+(1-i))}{B(\alpha,\beta)}\Big) \cdot
   \Big(\int_{M} \frac{x^{\alpha-1}(1-x)^{\beta-1}}
      {B(\alpha+i,\beta+(1-i))} \cdot x^{i} \cdot (1-x)^{1-i} \intd x\Big)
\\[+1em]
& = &
\displaystyle \int_{M}\frac{x^{\alpha-1}(1-x)^{\beta-1}}{B(\alpha,\beta)} 
   \cdot x^{i} \cdot (1-x)^{1-i} \intd x
\\[+1em]
& = &
\displaystyle \int_{M} u(\alpha,\beta,x)\cdot v(x, i) \intd x.
\end{array} \]

\noindent The latter expression is the left-hand-side
of~\eqref{eqn:conjugatepriorfun}. We see that the essence of the
verification of the conjugate prior equation is the shifting of
functions and normalisation factors.  This is a general pattern.

\auxproof{
We have to prove Equation~\eqref{eqn:conjugateprior}. Its
left-hand-side is, as Kleisli map $\NNO_{>0}\times\NNO_{>0}
\rightarrow \Giry([0,1]\times 2)$, applied to measurable subsets
$M\subseteq [0,1]$ and $N\subseteq 2$,
\[ \begin{array}{rcl}
\big(\tuple{\idmap, \flipchan} \after \betachan\big)(\alpha,\beta)(M\times N)
& = &
\big(\tuple{\idmap, \flipchan} \gg \betachan(\alpha,\beta)\big)(M\times N)
\\
& \smash{\stackrel{\eqref{eqn:pdfgraphstatetransformation}}{=}} &
\displaystyle \int_{M\times N} \frac{x^{\alpha-1}(1-x)^{\beta-1}}
   {B(\alpha,\beta)} \cdot x^{i}\cdot (1-x)^{1-i} \intd (x,i)
\\
& = &
\displaystyle \int_{M\times N} \frac{x^{\alpha-1+i}(1-x)^{\beta-i}}
   {B(\alpha,\beta)} \intd (x,i).
\end{array} \]

\noindent With a bit more effort we show that the right-hand-side
of~\eqref{eqn:conjugateprior} has the same outcome:
\[ \begin{array}{rcl}
\lefteqn{\big(((\betachan \after h)\otimes\idmap) \after
   \tuple{\idmap, \copychan \after \flipchan \after \betachan}\big)
   (\alpha, \beta)(M\times N)}
\\
& \smash{\stackrel{\eqref{eqn:graphequation}}{=}} &
\displaystyle \int ((\betachan \after h)\otimes\idmap)(-)(M\times N)
   \intd \big(\eta(\alpha,\beta) \otimes 
    (\copychan \after \flipchan \after \betachan)(\alpha,\beta)\big)
\\
& = &
\displaystyle \int ((\betachan \after h)\otimes\idmap)(\alpha, \beta, -)
   (M\times N)   \intd 
   \Giry(\copychan)\big((\flipchan \after \betachan)(\alpha,\beta)\big)
\\
& \smash{\stackrel{\eqref{??}}{=}} &
\displaystyle \int ((\betachan \after h)\otimes\idmap)(\alpha, \beta, 
   \copychan(-))(M\times N)   \intd 
   \big(\flipchan \after \betachan)(\alpha,\beta)\big)
\\
& \smash{\stackrel{(\ref{eqn:pdfintegration},
     \ref{eqn:pdfseqcomposition})}{=}} &
\displaystyle \int \big(\int \frac{x^{\alpha-1+i}(1-x)^{\beta-i}}
   {B(\alpha,\beta)}\intd x\big) \cdot
   ((\betachan \after h)\otimes\idmap)(\alpha, \beta, 
   \copychan(i))(M\times N) \intd i
\\
& = &
\displaystyle \int \frac{\int x^{\alpha-1+i}(1-x)^{\beta-i} \intd x}
   {B(\alpha,\beta)} \cdot
   \betachan(h(\alpha, \beta, i))(M) \cdot \indic{N}(i) \intd i
\\
& = &
\displaystyle \int_{N} \frac{B(\alpha+i, \beta+1-i)}{B(\alpha,\beta)} \cdot
   \betachan(\alpha+i, \beta+1-i)(M) \intd i
\\
& = &
\displaystyle \int_{N} \frac{B(\alpha+i, \beta+1-i)}{B(\alpha,\beta)} \cdot
   \big(\int_{M} \frac{x^{\alpha+i-1}\cdot (1-x)^{\beta-i}}{B(\alpha+i, \beta+1-i)}
   \intd x\big) \intd i
\\
& = &
\displaystyle \int_{M\times N} \frac{x^{\alpha-1+i}(1-x)^{\beta-i}}
   {B(\alpha,\beta)} \intd (x,i).
\end{array} \]
}
\end{example}

\begin{example}
\label{ex:betabinom}
In a similar way one verifies that the $\betachan$ channel is a
conjugate prior to the binomial channel. For the latter we fix a
natural number $n>0$, and consider the two channels:
\[ \xymatrix@-1pc{
\NNO_{>0}\times\NNO_{>0}\ar[rr]^-{\betachan} & &
   [0,1]\ar[rrr]^-{\binomchan_{n}} & & & \{0,1,\ldots,n\}
} \]

\noindent The binomial channel $\binomchan_{n}$ is defined for $r \in
          [0,1]$ and $M\subseteq \{0,1,\ldots,n\}$ as:
\[ \begin{array}{rcccl}
\binomchan_{n}(r)(M)
& = &
{\displaystyle \int_{M}} \binom{n}{i}\cdot r^{i}\cdot (1-r)^{n-i}
   \intd i
& = &
{\displaystyle \sum_{i\in M}}\, \binom{n}{i}\cdot r^{i}\cdot (1-r)^{n-i}.
\end{array} \]

\noindent The conjugate prior property requires in this situation a
parameter translation function $h\colon \NNO_{>0}\times\NNO_{>0}
\times \{0,1,\ldots,n\}\rightarrow \NNO_{>0}\times\NNO_{>0}$, which is
given by:
\[ \begin{array}{rcl}
h(\alpha, \beta, i)
& = &
(\alpha+i, \beta+n-i).
\end{array} \]

\noindent The proof of Equation~\eqref{eqn:conjugatepriorfun} is much
like in Example~\ref{ex:betaflip}, with $1-i$ replaced by $n-i$, and
an additional binomial term $\binom{n}{i}$ that is shifted from one
integral to another.

\begin{Auxproof}
\[ \begin{array}{rcl}
\lefteqn{\displaystyle\Big(\int \frac{x^{\alpha-1}(1-x)^{\beta-1}}{B(\alpha,\beta)}
   \cdot \binom{n}{i}\cdot x^{i} \cdot (1-x)^{n-i} \intd x\Big) \cdot
   \Big(\int_{M} \frac{x^{\alpha+i-1}(1-x)^{\beta+(n-i)-1}}
      {B(\alpha+i,\beta+(n-i))} \intd x\Big)}
\\[+1em]
& = &
\displaystyle\Big(\frac{\int x^{\alpha+i-1}(1-x)^{\beta+(n-i)-1} \intd x}
      {B(\alpha,\beta)}\Big) \cdot \binom{n}{i} \cdot
   \Big(\int_{M} \frac{x^{\alpha-1}(1-x)^{\beta-1}}
      {B(\alpha+i,\beta+(n-i))} \cdot x^{i} \cdot (1-x)^{n-i} \intd x\Big)
\\
& = &
\displaystyle\Big(\frac{B(\alpha+i,\beta+(n-i))}{B(\alpha,\beta)}\Big) \cdot
   \Big(\int_{M} \frac{x^{\alpha-1}(1-x)^{\beta-1}}{B(\alpha+i,\beta+(n-i))} \cdot 
   \binom{n}{i}\cdot x^{i} \cdot (1-x)^{n-i} \intd x\Big)
\\
& = &
\displaystyle \int_{M}\frac{x^{\alpha-1}(1-x)^{\beta-1}}{B(\alpha,\beta)} 
   \cdot \binom{n}{i} \cdot x^{i} \cdot (1-x)^{n-i} \intd x.
\end{array} \]
\end{Auxproof}
\end{example}

Here is another well-known conjugate prior relationship, namely
between Dirichlet and `multinomial' distributions. The latter
are simply called discrete distributions in the present context.

\begin{example}
\label{ex:dirmon}
Here we shall identify a number $n\in\NNO$ with the $n$-element set
$\{0,1,\ldots,n-1\}$. We then write $\Dst_{*}(n)$ for the set of
$n$-tuples $(x_{0}, \ldots, x_{n-1})\in (\R_{>0})^{n}$ with
$\sum_{i}x_{i} = 1$.

For a fixed $n>0$, let $O = \{y_{0}, \ldots,
y_{n-1}\}$ be a set of `observations'. We consider the following two
channels.
\[ \xymatrix@-1pc{
(\NNO_{>0})^{n}\ar[rr]^-{\dirchan_n} & & \Dst_{*}(n)\ar[rr]^-{\multchan} & & O
} \]

\noindent The multinomial channel is defined as $\multchan(x_{0},
\ldots, x_{n-1}) = x_{0}\ket{y_{0}} + \cdots + x_{n-1}\ket{y_{n-1}}$.
It can be described as a pdf-channel, via the function $v(\vec{x},y)
\coloneqq x_{i} \mbox{ if }y=y_{i}$. Then, for $N\subseteq O =
\{y_{0}, \ldots, y_{n-1}\}$,
\[ \begin{array}{rcccl}
\multchan(\vec{x})(N)
& = &
\displaystyle\int_{N} v(\vec{x}, y) \intd y
& = &
\sum\set{x_{i}}{y_{i}\in N}.
\end{array} \]

The Dirichlet channel $\dirchan_n$ is more complicated: for an
$n$-tuple $\vec{\alpha} = (\alpha_{0}, \ldots, \alpha_{n-1})$ it is
given via pdf's $d_n$, in:
\[ \begin{array}{rclcrcl}
\dirchan_{n}(\vec{\alpha})
& = &
\displaystyle\int d_{n}(\vec{\alpha})
& \qquad\mbox{where}\qquad &
d_{n}(\vec{\alpha})(x_{0}, \ldots, x_{n-1})
& = &
{\displaystyle\frac{\Gamma(\sum_{i}\alpha_{i})}{\prod_{i}\Gamma(\alpha_{i})}}
   \cdot \prod_{i} x_{i}^{\alpha_{i}-1},
\end{array} \]

\noindent for $(x_{0}, \ldots, x_{n-1}) \in \Dst_{*}(n)$. The
operation $\Gamma$ is the `Gamma' function, which is defined on
natural numbers $k > 1$ as $\Gamma(k) = (k-1)!$. Hence $\Gamma$ can be
defined recursively as $\Gamma(1) = 1$ and $\Gamma(k+1) =
k\cdot\Gamma(k)$. The above fraction is a normalisation factor since
one has $\frac{\prod_{i}
  \Gamma(\alpha_{i})}{\Gamma(\sum_{i}\alpha_{i})} = \int\prod_{i}
x_{i}^{\alpha_{i}-1} \intd \vec{x}$, see
\textit{e.g.}~\cite{Bishop06}. From this one can derive: $\int
x_{i}\cdot d_{n}(\vec{\alpha})(\vec{x}) \intd \vec{x} =
\frac{\alpha_{i}}{\sum_{j}\alpha_j}$.

The parameter translation function $h\colon (\NNO_{>0})^{n} \times O
\rightarrow (\NNO_{>0})^{n}$ is:
\[ \begin{array}{rcl}
h(\alpha_{0}, \ldots, \alpha_{n-1}, y)
& = &
(\alpha_{0}, \ldots, \alpha_{i}+1, \ldots, \alpha_{n-1})
\quad \mbox{if } y=y_{i}.
\end{array} \]

\noindent We check Equation~\eqref{eqn:conjugatepriorfun}, for $M\subseteq
\Dst_{*}(n)$ and observation $y_{i}\in O$,
\[ \begin{array}{rcl}
\lefteqn{\displaystyle
  \Big(\int d_{n}(\vec{\alpha})(\vec{x})\cdot v(\vec{x},y_{i})\intd\vec{x}\Big)
   \cdot \Big(\int_{M} d_{n}(h(\vec{\alpha},y_{i}))(\vec{x}) \intd \vec{x}\Big)}
\\[+0.4em]
& = &
\frac{\alpha_i}{\sum_{j}\alpha_{j}} \cdot 
{\displaystyle\int_{M}} 
   \frac{\Gamma(1 + \sum_{j}\alpha_{j})}
   {\Gamma(\alpha_{i}+1)\cdot \prod_{j\neq i}\Gamma(\alpha_{j})} \cdot
   x_{i}^{\alpha_i} \cdot \prod_{j\neq i} x_{j}^{\alpha_{j}-1} \intd \vec{x}
\\[+0.8em]
& = &
{\displaystyle\Big(\int_{M}} 
   \frac{\Gamma(\sum_{j}\alpha_{j})}
   {\prod_{j}\Gamma(\alpha_{j})} \cdot
   x_{i} \cdot \prod_{j} x_{j}^{\alpha_{j}-1} \intd \vec{x}
\\[+0.8em]
& = &
\displaystyle\int_{M} d_{n}(\vec{\alpha})(\vec{x})\cdot 
   v(\vec{x},y_{i})\intd\vec{x}.
\end{array} \]

\end{example}

% Distinctions of Schroeder:
% - unknown mean, known variance
% - known mean, unknown variance
% - unknown mean, unknown variance

We include one more example, illustrating that normal channels are
conjugate priors to themselves. This fact is also well-known. The
point is to illustrate once again how that works in the current
setting.

\begin{example}
\label{ex:normnorm}
Consider the following two normal channels.
\[ \xymatrix@-0.5pc{
\R\times\R_{>0}\ar[rr]^-{\normchan} & &
   \R\ar[rrr]^-{\normchan(-,\nu)} & & & \R_{>0}
} \]

\noindent The channel $\normchan$ is described explicitly
in~\eqref{diag:normal}. Notice that we use it twice here, the second
time with a fixed standard deviation $\nu$, for `noise'. This second
channel is typically used for observation, like in Kalman filtering,
for which a fixed noise level can be assumed. We claim that the first
normal channel $\normchan$ is a conjugate prior to the second channel
$\normchan(-,\nu)$, via the parameter translation function $h\colon
\R\times\R_{>0}\times\R_{>0}\rightarrow \R\times\R_{>0}$ given by:
\[ \begin{array}{rcl}
h(\mu, \sigma, y)
& = &
\displaystyle
(\, \frac{\mu\cdot\nu^{2} + y\cdot\sigma^{2}}{\nu^{2}+\sigma^{2}},
   \frac{\nu\cdot\sigma}{\sqrt{\nu^{2}+\sigma^{2}}} \,)
\end{array} \]

% Note: the new standard deviation does not depend on y.

\noindent We prove Equation~\eqref{eqn:conjugatepriorfun}, again
starting from the right.
\[ \begin{array}{rcl}
\lefteqn{\Big({\displaystyle \int}
  \frac{1}{\sqrt{2\pi}\sigma}e^{-\frac{(x-\mu)^{2}}{2\sigma^{2}}} \cdot
  \frac{1}{\sqrt{2\pi}\nu}e^{-\frac{(y-x)^{2}}{2\nu^{2}}}\intd x\Big) \cdot
  \Big({\displaystyle \int_{M}} 
  \frac{\sqrt{\nu^{2}+\sigma^{2}}}{\sqrt{2\pi}\nu\sigma}
   e^{-\frac{(\nu^{2}+\sigma^{2}) (x-\frac{\mu\cdot\nu^{2} + y\cdot\sigma^{2}}
    {\nu^{2}+\sigma^{2}})^{2}}{2\nu^{2}\sigma^{2}}}\intd x\Big)}
\\[+1em]
& \smash{\stackrel{(*)}{=}} &
\Big({\displaystyle \int}\frac{1}{\sqrt{2\pi}\sigma\nu}
  e^{-\frac{\nu^{2}(x-\mu)^{2} + \sigma^{2}(y-x)^{2}}{2\sigma^{2}\nu^{2}}}\intd x\Big) 
\\
& & \qquad \cdot\; \Big({\displaystyle \int_{M}} 
  \frac{\sqrt{\nu^{2}+\sigma^{2}}}{\sqrt{2\pi}\nu\sigma}
   e^{-\frac{\nu^{2}(x-\mu)^{2} + \sigma^{2}(y-x)^{2} - \nu^{2}\mu^{2} - 
      \sigma^{2}y^{2} + \frac{(\mu\cdot\nu^{2} + y\cdot\sigma^{2})^{2}}
    {\nu^{2}+\sigma^{2}}}{2\nu^{2}\sigma^{2}}}\intd x\Big)
\\
& = &
\Big({\displaystyle \int}\frac{\sqrt{\nu^{2}+\sigma^{2}}}{\sqrt{2\pi}\nu\sigma}
  e^{-\frac{\nu^{2}(x-\mu)^{2} + \sigma^{2}(y-x)^{2} - \nu^{2}\mu^{2} - 
      \sigma^{2}y^{2} + \frac{(\mu\cdot\nu^{2} + y\cdot\sigma^{2})^{2}}
    {\nu^{2}+\sigma^{2}}}{2\sigma^{2}\nu^{2}}}\intd x\Big) 
\\
& & \qquad \cdot\; \Big({\displaystyle \int_{M}} 
  \frac{1}{\sqrt{2\pi}\sigma\nu}
   e^{-\frac{\nu^{2}(x-\mu)^{2} + \sigma^{2}(y-x)^{2}}{2\nu^{2}\sigma^{2}}}\intd x\Big)
\\
& \smash{\stackrel{(*)}{=}} &
\Big({\displaystyle \int} 
  \frac{\sqrt{\nu^{2}+\sigma^{2}}}{\sqrt{2\pi}\nu\sigma}
   e^{-\frac{(\nu^{2}+\sigma^{2}) (x-\frac{\mu\cdot\nu^{2} + y\cdot\sigma^{2}}
    {\nu^{2}+\sigma^{2}})^{2}}{2\nu^{2}\sigma^{2}}}\intd x\Big) \cdot
   \Big({\displaystyle \int_{M}}
   \frac{1}{\sqrt{2\pi}\sigma}e^{-\frac{(x-\mu)^{2}}{2\sigma^{2}}} \cdot
   \frac{1}{\sqrt{2\pi}\nu}e^{-\frac{(y-x)^{2}}{2\nu^{2}}}\intd x\Big)
\\[+1em]
& = &
{\displaystyle \int_{M}}
  \frac{1}{\sqrt{2\pi}\sigma}e^{-\frac{(x-\mu)^{2}}{2\sigma^{2}}} \cdot
  \frac{1}{\sqrt{2\pi}\nu}e^{-\frac{(y-x)^{2}}{2\nu^{2}}}\intd x.
\end{array} \]

\noindent The last equation holds because the first integral in the
previous line equals one, since, in general, the integral over a pdf
is one. The two marked equations $\smash{\stackrel{(*)}{=}}$ are
justified by:
\[ \begin{array}{rcl}
\lefteqn{(\nu^{2}+\sigma^{2})\big(x-\frac{\mu\cdot\nu^{2} + 
   y\cdot\sigma^{2}}{\nu^{2}+\sigma^{2}}\big)^{2}}
\\
& = &
(\nu^{2}+\sigma^{2})x^{2}-
   2(\mu\cdot\nu^{2} + y\cdot\sigma^{2})x
    + \frac{(\mu\cdot\nu^{2} + y\cdot\sigma^{2})^{2}}
    {\nu^{2}+\sigma^{2}}
\\
& = &
\nu^{2}(x^{2}-2\mu x + \mu^{2}) + \sigma^{2}(y^{2}-2yx + x^{2})
    - \nu^{2}\mu^{2} - \sigma^{2}y^{2}
    + \frac{(\mu\cdot\nu^{2} + y\cdot\sigma^{2})^{2}}
    {\nu^{2}+\sigma^{2}}
\\
& = &
\nu^{2}(x-\mu)^{2} + \sigma^{2}(y-x)^{2} 
    - \nu^{2}\mu^{2} - \sigma^{2}y^{2}
    + \frac{(\mu\cdot\nu^{2} + y\cdot\sigma^{2})^{2}}
    {\nu^{2}+\sigma^{2}}
\end{array} \]
\end{example}

\section{Conjugate priors form Bayesian inversions}\label{sec:conjugatepriorinversion}

This section connects the main two notions of this paper, by showing
that conjugate priors give rise to Bayesian inversion. The argument is
a very simple example of diagrammatic reasoning. Before we come to
it, we have to clarify an issue that was left open earlier, regarding
`deterministic' channels, see Definition~\ref{def:conjugateprior}.

\begin{definition}
\label{def:deterministic}
A channel $c$ is called \emph{deterministic} if it commutes with
copiers, that is, if it satisfies the equation on the left below.
\[ \vcenter{\hbox{%
\begin{tikzpicture}[font=\small]
\node[copier] (c) at (0,0.6) {};
\node[arrow box] (f) at (0,0.1) {$c$};
\draw (c) to[out=15,in=-90] (0.4,1.0);
\draw (c) to[out=165,in=-90] (-0.4,1.0);
\draw (f) to (c);
\draw (f) to (0,-0.4);
\end{tikzpicture}}}
\quad=\quad
\vcenter{\hbox{%
\begin{tikzpicture}[font=\small]
\node[copier] (c) at (0,0.0) {};
\node[arrow box] (f1) at (-0.4,0.6) {$c$};
\node[arrow box] (f2) at (0.4,0.6) {$c$};
\draw (c) to[out=165,in=-90] (f1);
\draw (c) to[out=15,in=-90] (f2);
\draw (f1) to (-0.4,1.1);
\draw (f2) to (0.4,1.1);
\draw (c) to (0,-0.3);
\end{tikzpicture}}}
\hspace*{10em}
\vcenter{\hbox{%
\begin{tikzpicture}[font=\small]
\node[copier] (c) at (0,0.4) {};
\node[state] (omega) at (0,0.1) {$\omega$};
\draw (c) to[out=15,in=-90] (0.4,0.8);
\draw (c) to[out=165,in=-90] (-0.4,0.8);
\draw (omega) to (c);
\end{tikzpicture}}}
\quad=\quad
\vcenter{\hbox{%
\begin{tikzpicture}[font=\small]
\node[state] (omega1) at (-0.5,0.0) {$\omega$};
\node[state] (omega2) at (0.5,0.0) {$\omega$};
\draw (omega1) to (-0.5,0.4);
\draw (omega2) to (0.5,0.4);
\end{tikzpicture}}}
\]

\noindent As a special case, a state $\omega$ is called deterministic
if it satisfies the equation on the right, above.
\end{definition}

The state description is a special case of the channel description
since a state on $X$ is a channel $1\rightarrow X$ and copying on the
trivial (final) object $1$ does nothing, up to isomorphism.

Few channels (or states) are deterministic. In deterministic and
continuous computation, the ordinary functions $f\colon X \rightarrow
Y$ are deterministic, when considered as a channel $\eta \after f$.
We check this explicitly for point states, since this is what we need
later on.

\begin{example}
\label{ex:deterministicstate}
Let $x$ be an element of a measurable space $X$. The associated point
state $\eta(x) \in \Giry(X)$ is deterministic, where $\eta(x)(M) =
\indic{M}(x)$. We check the equation on the right in
Definition~\ref{def:deterministic}:
\[ \begin{array}{rcl}
\big(\copier \after \eta(x)\big)(M\times N)
& = &
\eta(x,x)(M\times N)
\hspace*{\arraycolsep}=\hspace*{\arraycolsep}
\indic{M\times N}(x,x)
\hspace*{\arraycolsep}=\hspace*{\arraycolsep}
\indic{M}(x) \cdot \indic{N}(x)
\\
& = &
\eta(x)(M)\cdot \eta(x)(N)
\hspace*{\arraycolsep}=\hspace*{\arraycolsep}
\big(\eta(x)\otimes\eta(x)\big)(M\times N).
\end{array} \]
\end{example}

We now come to the main result.

\begin{theorem}
\label{thm:conjugateinversion}
Let $\smash{P \stackrel{c}{\rightarrow} X \stackrel{d}{\rightarrow}
  O}$ be channels, where $c$ is conjugate prior to $d$, say via
$h\colon P \times O \rightarrow P$. Then for each deterministic
(copyable) state $p$, the map $c \after h(p, -) \colon O \rightarrow
X$ is a Bayesian inversion of $d$, wrt.\ the transformed state $c \gg
p$.
\end{theorem}

\begin{myproof}
We have to prove Equation~\eqref{diag:inversion}, for channel $d$ and
state $c \gg p$, with the channel $c \after h(p, -)$ playing the role
of Bayesian inversion $d^{\dag}_{c \gg p}$. This is easiest to see
graphically, using that the state $p$ is deterministic and thus
commutes with copiers $\copier$, see the equation on the right in
Definition~\ref{def:deterministic}.
\[ \vcenter{\hbox{%
\begin{tikzpicture}[font=\small]
\node[state] (p) at (0,-0.7) {$p$};
\node[arrow box] (c) at (0.0,-0.2) {$c$};
\node[copier] (copier) at (0,0.3) {};
\node[arrow box] (d) at (0.5,0.95) {$d$};
\coordinate (X) at (-0.5,1.5);
\coordinate (Y) at (0.5,1.5);
\draw (c) to (copier);
\draw (c) to (p);
\draw (copier) to[out=150,in=-90] (X);
\draw (copier) to[out=15,in=-90] (d);
\draw (d) to (Y);
\end{tikzpicture}}}
\quad 
\smash{\stackrel{\eqref{eqn:conjugateprior}}{=}}
\quad
\vcenter{\hbox{%
\begin{tikzpicture}[font=\small]
\node[state] (p) at (0,0.0) {$p$};
\node[copier] (copier1) at (0,0.3) {};
\node[copier] (copier2) at (0.25,2.0) {};
\coordinate (X) at (-0.5,3.8);
\coordinate (Y) at (0.5,3.8);
\node[arrow box] (c1) at (0.25,0.8) {$c$};
\node[arrow box] (d) at (0.25,1.5) {$d$};
\node[arrow box] (h) at (-0.5,2.6) {$\;\;h\;\;$};
\node[arrow box] (c2) at (-0.5,3.3) {$c$};
\draw (copier1) to (p);
\draw (copier1) to[out=150,in=-90] (h.240);
\draw (copier1) to[out=30,in=-90] (c1);
\draw (c1) to (d);
\draw (d) to (copier2);
\draw (copier2) to[out=165,in=-90] (h.305);
\draw (h) to (c2);
\draw (c2) to (X);
\draw (copier2) to[out=30,in=-90] (Y);
\end{tikzpicture}}}
\quad
=
\quad
\vcenter{\hbox{%
\begin{tikzpicture}[font=\small]
\node[state] (p1) at (-0.65,2.1) {$p$};
\node[state] (p2) at (0.25,-0.15) {$p$};
\node[copier] (copier2) at (0.25,1.5) {};
\coordinate (X) at (-0.5,3.8);
\coordinate (Y) at (0.5,3.8);
\coordinate (Z) at (0.0,2.0);
\node[arrow box] (c1) at (0.25,0.3) {$c$};
\node[arrow box] (d) at (0.25,1.0) {$d$};
\node[arrow box] (h) at (-0.5,2.6) {$\;\;h\;\;$};
\node[arrow box] (c2) at (-0.5,3.3) {$c$};
\draw (p1) to[out=90,in=-90] (h.240);
\draw (p2) to[out=90,in=-90] (c1);
\draw (c1) to (d);
\draw (d) to (copier2);
\draw (copier2) to[out=165,in=-90] (Z);
\draw (Z) to[out=90,in=-90] (h.305);
\draw (h) to (c2);
\draw (c2) to (X);
\draw (copier2) to[out=30,in=-90] (Y);
\end{tikzpicture}}} \]

\noindent This is it. \QED
\end{myproof}

When we specialise to Giry-channels we get an `if-and-only-if'
statement, since there we can reason elementwise.

\begin{corollary}
\label{cor:Giry}
Let $\smash{P\xrightarrow{c} X\xrightarrow{d} O}$ be two channels in
$\Kl(\Giry)$, and let $h \colon P\times O \rightarrow P$ be a
measurable function. The following two points are equivalent:
\begin{itemize}
\item[(i)] $c$ is a conjugate prior to $d$, via $h$;

\item[(ii)] $c(h(p,-)) \colon O \rightarrow \Giry(X)$ is a Bayesian
  inverse for channel $d$ with state $c(p)$, \textit{i.e.}~is
  $d^{\dag}_{c(p)}$, for each parameter $p\in P$. \QED
\end{itemize}
\end{corollary}

\section{A logical perspective on conjugate priors}\label{sec:logic}

This section takes a logically oriented, look at conjugate priors,
describing them in terms of updates of a prior state with a random
variable (or predicate). This new perspective is interesting for two
reasons:
\begin{itemize}
\item it formalises the intuition behind conjugate priors in a
  precise manner, see
  \textit{e.g.}\ Equations~\eqref{eqn:betaflipupdateeqn}
  and~\eqref{eqn:betabinomupdateeqn} below, where the characteristic
  closure property for a class of distributions is expressed via
  occurrences of these distributions on both sides of an equation;

\item it will be useful in the next section to capture multiple
  observations via an update with a conjunction of multiple random
  variables.
\end{itemize}

\noindent But first we need to introduce some new terminology. We
shall do so separately for discrete and continuous probability,
although both can be described as instances of the same category
theoretic notions, using effectus theory~\cite{Jacobs15d,Jacobs17a}.

\subsection{Discrete updating}

A \emph{random variable} on a set $X$ is a function $r\colon X
\rightarrow \R$. It is a called a \emph{predicate} if it restricts to
$X\rightarrow [0,1]$. Simple examples of predicates are indicator
functions $\indic{E} \colon X \rightarrow [0,1]$, for a subset/event
$E\subseteq X$, given by $\indic{E}(x) = 1$ if $x\in E$ and
$\indic{E}(x) = 0$ if $x\not\in E$. Indicator functions $\indic{\{x\}}
\colon X \rightarrow [0,1]$ for a singleton subset are sometimes
called point predicates. For two random variables $r,s\colon X
\rightarrow \R$ we write $r\andthen s\colon X \rightarrow \R$ for the
new variable obtained by pointwise multiplication: $(r\andthen s)(x) =
r(x) \cdot s(x)$.

For a random variable $r$ and a discrete
probability distribution (or state) $\omega\in\Dst(X)$ we define the
\emph{validity} $\omega\models r$ as the expected value:
\begin{equation}
\label{eqn:discvalidity}
\begin{array}{rcl}
\omega\models r
& \;\coloneqq\; &
\displaystyle\sum_{x\in X} \omega(x)\cdot r(x).
\end{array}
\end{equation}

\noindent Notice that this is a finite sum, since by definition the
support of $\omega$ is finite.

If we have a channel $c\colon X \rightarrow \Dst(Y)$ and a random
variable $r\colon Y\rightarrow \R$ on its codomain $Y$, then we can
transform it --- or pull it back --- into a random variable on its
domain $X$. We write this pulled back random variable as $c \ll r
\colon X \rightarrow \R$. It is defined as:
\begin{equation}
\label{eqn:discrandvartransform}
\begin{array}{rcccl}
\big(c \ll r\big)(x)
& \coloneqq &
c(x) \models r
& = &
\displaystyle\sum_{y\in Y} c(x)(y)\cdot r(y).
\end{array}
\end{equation}

\noindent This operation $\ll$ interacts nicely with composition
$\after$ of channels, in the sense that $(d\after c) \ll r = c \ll (d
\ll r)$. Moreover, the validity $\omega \models c \ll r$ is the same
as the validity $c\gg \omega\models r$, where $\gg$ is state
transformation, see~\eqref{eqn:discstatransf}.

If a validity $\omega\models r$ is non-zero, then we can define the
\emph{updated} or \emph{conditioned} state $\omega|_{r} \in \Dst(X)$
via:
\begin{equation}
\label{eqn:discconditioning}
\begin{array}{rclcrcl}
\big(\omega|_{r}\big)(x)
& \coloneqq &
\displaystyle\frac{\omega(x)\cdot r(x)}{\omega\models r}
& \qquad\mbox{that is}\qquad &
\omega|_{r}
& = &
\displaystyle\sum_{x\in X}
   \frac{\omega(x)\cdot r(x)}{\omega\models r}\bigket{x}.
\end{array}
\end{equation}

\noindent The first formulation describes the updated distribution
$\omega|_{r}$ as a probability mass function, whereas the second one
uses a formal convex sum.

It is not hard to see that successive updates commute and can be
reduced to a single update via $\andthen$, as in:
\begin{equation}
\label{eqn:discconditioningand}
\begin{array}{rcccccl}
\big(\omega|_{r}\big)|_{s}
& = &
\omega|_{r\andthen s}
& = &
\omega|_{s\andthen r}
& = &
\big(\omega|_{s}\big)|_{r}.
\end{array}
\end{equation}

\noindent One can multiply a random variable $r\colon X \rightarrow
\R$ with a scalar $a\in\R$, pointwise, giving a new random variable
$a\cdot r \colon X \rightarrow \R$. When $a\neq 0$ it disappears from
updating:
\begin{equation}
\label{eqn:discconditioningscal}
\begin{array}{rcl}
\omega|_{a\cdot r}
& = &
\omega|_{r}.
\end{array}
\end{equation}

\begin{proposition}
\label{prop:conjpriorpointupdate}
Assume that composable channels $\smash{P \stackrel{c}{\rightarrow} X
  \stackrel{d}{\rightarrow} O}$ for the discrete distribution monad
$\Dst$ are given, where $c$ is conjugate prior to $d$, say via
$h\colon P \times O \rightarrow P$. The distribution for the updated
parameter $h(p,y)$ is then an update of the distribution for the
original parameter $p$, with the pulled-back point predicate for the
observation $y$, as in:
\[ \begin{array}{rcl}
c\big(h(p,y)\big)
& = &
c(p)\big|_{d \ll \indic{\{y\}}}.
\end{array} \]
\end{proposition}

\begin{myproof}
We first notice that the pulled-back singleton predicate $d \ll
\indic{\{y\}} \colon X \rightarrow \R$ is:
\[ \begin{array}{rcccl}
(d \ll \indic{\{y\}})(x)
& \smash{\stackrel{\eqref{eqn:discrandvartransform}}{=}} &
\displaystyle\sum_{z\in Y} d(x)(z) \cdot \indic{\{y\}}(z)
& = &
d(x)(y).
\end{array} \]

\noindent Theorem~\ref{thm:conjugateinversion} tells us that
$c\big(h(p,y)\big)$ is obtained via the Bayesian inversion of $d$, so
that:
\[ \begin{array}[b]{rcl}
c\big(h(p,y)\big)(x)
& = &
d^{\dag}_{c(p)}(y)(x) 
\\[+0.4em]
& \smash{\stackrel{\eqref{eqn:discreteinversion}}{=}} &
\displaystyle\frac{c(p)(x)\cdot d(x)(y)}{\sum_{z} c(p)(z)\cdot d(z)(y)}
\\[+1em]
& = &
\displaystyle\frac{c(p)(x)\cdot (d \ll \indic{\{y\}})(x)}
   {\sum_{z} c(p)(z)\cdot (d \ll \indic{\{y\}})(z)}
   \qquad \mbox{as just noted}
\\[+1em]
& = &
\displaystyle\frac{c(p)(x)\cdot (d \ll \indic{\{y\}})(x)}
   {c(p) \models d \ll \indic{\{y\}}}
\\
& \smash{\stackrel{\eqref{eqn:discconditioning}}{=}} &
c(p)\big|_{d \ll \indic{\{y\}}}.
\end{array} \eqno{\QEDbox} \]
\end{myproof}

In fact, what we are using here is that the Bayesian inversion
$c^{\dag}_{\omega}$ defined in~\eqref{eqn:discreteinversion} is an
update: $c^{\dag}_{\omega} = \omega|_{c \ll \indic{\{y\}}}$.

\subsection{Continuous updating}

We now present the analogous story for continuous probability.  A
\emph{random variable} on a measurable space $X$ is a measurable
function $X\rightarrow \R$. It is called a \emph{predicate} if it
restricts to $X\rightarrow [0,1]$. These random variables (and
predicates) are closed under $\andthen$ and scalar multiplication,
defined via pointwise multiplication. In the continuous case one
typically has no point predicates.

Given a measure/state $\omega\in\Giry(X)$ and a random variable
$r\colon X \rightarrow \R$ we define the validity $\omega\models r$
again as expected value:
\begin{equation}
\label{eqn:contvalidity}
\begin{array}{rcl}
\omega\models r
& \;\coloneqq\; &
\displaystyle\int r \intd\omega.
\end{array}
\end{equation}

\noindent This allows us to define transformation of a random
variable, backwards along a channel: for a channel $c\colon
X\rightarrow \Giry(Y)$ and a random variable $r\colon Y\rightarrow \R$
we write $c \ll r \colon X \rightarrow \R$ for the pulled-back random
variable defined by:
\begin{equation}
\label{eqn:contrandvartransform}
\begin{array}{rcccl}
\big(c \ll r\big)(x)
& \coloneqq &
c(x) \models r
& = &
\displaystyle\int r \intd\, c(x).
\end{array}
\end{equation}

\noindent The update $\omega|_{r}\in\Giry(X)$ of a state
$\omega\in\Giry(X)$ with a random variable $r\colon X \rightarrow \R$
is defined on a measurable subset $M\subseteq X$ as:
\begin{equation}
\label{eqn:contconditioning}
\begin{array}{rcccccl}
\big(\omega|_{r}\big)(M)
& \coloneqq &
\displaystyle\frac{\int_{M} r \intd\omega}{\int r \intd\omega}
& = &
\displaystyle\frac{\int_{M} r \intd\omega}{\omega\models r}
& = &
\displaystyle\int_{M}\frac{r}{\omega\models r} \intd\omega.
\end{array}
\end{equation}

\noindent If $\omega = \int f$ for a pdf $f$, this becomes:
\begin{equation}
\label{eqn:contconditioningpdf}
\begin{array}{rcccccl}
\big(\omega|_{r}\big)(M)
& \coloneqq &
\displaystyle\frac{\int_{M} f(x) \cdot r(x) \intd x}
   {\int f(x) \cdot r(x) \intd x}
& = &
\displaystyle\frac{\int_{M} f(x) \cdot r(x) \intd x}{\omega\models r}
& = &
\displaystyle\int_{M} \frac{f(x) \cdot r(x)}{\omega\models r} \intd x.
\end{array}
\end{equation}

\noindent The latter formulation shows that the pdf of $\omega|_{r}$
is the function $x \mapsto \frac{f(x) \cdot r(x)}{\omega\models r}$.
Updating in the continuous case also satisfies the multiple-update and
scalar properties~\eqref{eqn:discconditioningand} and
~\eqref{eqn:discconditioningscal}.

Again we redescribe conjugate priors in terms of updating.

\begin{proposition}
\label{prop:contpriorpointupdate}
Let $\smash{P \stackrel{c}{\rightarrow} X \stackrel{d}{\rightarrow}
  O}$ be channels for the Giry monad $\Giry$, where $c$ and $d$ are
pdf-channels $c = \int u$ and $d = \int v$, for $u\colon P\times X
\rightarrow \R_{\geq 0}$ and $v\colon X\times O \rightarrow \R_{\geq
  0}$.  Assume that $c$ be conjugate prior to $d$ via $h\colon P
\times O \rightarrow P$. Then:
\[ \begin{array}{rcl}
c\big(h(p,y)\big)
& = &
c(p)\big|_{v(-,y)},
\end{array} \]

\noindent where $v(-,y) \colon O \rightarrow \R$ is used as random
variable on $O$.
\end{proposition}

\begin{myproof}
Theorem~\ref{thm:conjugateinversion} gives the first step in:
\[ \begin{array}[b]{rcl}
c\big(h(p,y)\big)(M)
& = &
d^{\dag}_{c(p)}(y)(M)
\\[+0.4em]
& \smash{\stackrel{\eqref{eqn:continuousinversion}}{=}} &
\displaystyle\frac{\int_{M} u(p,x)\cdot v(x,y) \intd x}
   {\int u(p,x) \cdot v(x,y) \intd x}
\\[+1em]
& = &
\displaystyle\frac{\int_{M} u(p,x)\cdot v(x,y) \intd x}
   {c(p) \models v(-,y)}
\\
& \smash{\stackrel{\eqref{eqn:contconditioningpdf}}{=}} &
c(p)\big|_{v(-,y)}(M).
\end{array} \eqno{\QEDbox} \]
\end{myproof}

The previous two propositions deal with two \emph{discrete} channels
$c,d$ (for $\Dst$) or with two \emph{continuous} channels (for
$\Giry$). But the update approach also works for mixed channels,
technically because $\Dst$ is a submonad of $\Giry$. We shall not
elaborate these details but give illustrations instead.

\begin{example}
\label{ex:betaflipupdate}
We shall have another look at the $\betachan - \flipchan$ conjugate
prior situation from Example~\ref{ex:betaflip}. We claim that the
essence of these channels being conjugate prior, via the parameter
translation function~\eqref{eqn:betaflipfun}, can be expressed via the
following two state update equations:
\begin{equation}
\label{eqn:betaflipupdateeqn}
\begin{array}{rcl}
\betachan(\alpha,\beta)\big|_{\flipchan \ll \indic{\{1\}}}
& = &
\betachan(\alpha+1,\beta)
\\
\betachan(\alpha,\beta)\big|_{\flipchan \ll \indic{\{0\}}}
& = &
\betachan(\alpha,\beta+1).
\end{array}
\end{equation}

\noindent These equations follow from what we have proven above. But
we choose to re-prove them here in order to illustrate how updating
works concretely. First note that for a parameter $x\in [0,1]$ we have
predicate values $\big(\flipchan \ll \indic{\{1\}}\big)(x) = x$ and
$\big(\flipchan \ll \indic{\{0\}}\big)(x) = 1-x$. Then:
\[ \begin{array}{rcl}
\betachan(\alpha,\beta) \models \flipchan \ll \indic{\{1\}}
\hspace*{\arraycolsep}\smash{\stackrel{\eqref{eqn:beta}}{=}}\hspace*{\arraycolsep}
\displaystyle\int x\cdot \frac{x^{\alpha-1}(1-x)^{\beta-1}}{B(\alpha,\beta)}\intd x
& = &
\displaystyle\frac{\int x^{\alpha}(1-x)^{\beta-1}\intd x}{B(\alpha,\beta)}
\\[+0.8em]
& = &
\displaystyle\frac{B(\alpha+1,\beta)}{B(\alpha,\beta)}.
\end{array} \]

\noindent Thus, using~\eqref{eqn:contconditioningpdf}, we obtain
the first equation in~\eqref{eqn:betaflipupdateeqn}:
\[ \begin{array}{rcl}
\betachan(\alpha,\beta)\big|_{\flipchan \ll \indic{\{1\}}}(M)
& = &
\displaystyle\frac{B(\alpha,\beta)}{B(\alpha+1,\beta)} \cdot
   \int_{M} x\cdot \frac{x^{\alpha-1}(1-x)^{\beta-1}}{B(\alpha,\beta)}\intd x
\\[+0.8em]
& = &
\displaystyle\int_{M} \frac{x^{\alpha}(1-x)^{\beta-1}}{B(\alpha+1,\beta)}\intd x
\\
& = &
\betachan(\alpha+1,\beta)(M).
\end{array} \]

In a similar way we can capture the $\betachan - \binomchan$ conjugate
priorship from Example~\ref{ex:betabinom} as update equation:
\begin{equation}
\label{eqn:betabinomupdateeqn}
\begin{array}{rcl}
\betachan(\alpha,\beta)\big|_{\binomchan_{n} \ll \indic{\{i\}}}
& = &
\betachan(\alpha+i, \beta+n-i).
\end{array}
\end{equation}

\noindent This equation, and also~\eqref{eqn:betaflipupdateeqn},
hightlight the original ideal behind conjugate priors, expressed
informally in many places in the literature as: we have a class of
distributions --- $\betachan$ in this case --- which is closed under
updates in a particular statistiscal model --- $\flipchan$ or
$\binomchan$ in these cases.
\end{example}

These update formulations~\eqref{eqn:betabinomupdateeqn}
and~\eqref{eqn:betaflipupdateeqn} may be useful when trying to find a
parameter translation function: one can start calculating the state
update on the left-hand-side, using
formulas~\eqref{eqn:discconditioning}
and~\eqref{eqn:contconditioning}, hoping that a distribution of the
same form appears (but with different parameters).

\section{Multiple updates}\label{sec:multiple}

So far we have dealt with the situation where there is a single
observation $y\in O$ that leads to an update of a prior
distribution. In this final section we briefly look at how to handle
multiple observations $y_{1}, \ldots, y_{m}$. This is what typically
happens in practice; it will lead to the notion of \emph{sufficient
  statistic}.

A good starting point is the $\betachan - \flipchan$ relationship from
Example~\ref{ex:betaflip} and~\ref{ex:betaflipupdate}, especially in
its snappy update form~\eqref{eqn:betaflipupdateeqn}. Suppose we have
multiple head/tail observations $y_{1}, \ldots, y_{m} \in 2 = \{0,1\}$
which we wish to incorporate into a prior distribution
$\betachan(\alpha,\beta)$. Following
Equation~\eqref{eqn:betaflipupdateeqn} we use multiple updates, on the
left below, which can be rewritten as a single update, on the
right-hand-side of the equation via conjunction $\andthen$,
using~\eqref{eqn:discconditioningand}:
\[ \begin{array}{rcl}
\lefteqn{\betachan(\alpha,\beta)\big|_{\flipchan \ll \indic{\{y_1\}}} 
  \big|_{\flipchan \ll \indic{\{y_2\}}} \;\cdots\; \big|_{\flipchan \ll \indic{\{y_m\}}}}
\\
& = &
\betachan(\alpha,\beta)\big|_{(\flipchan \ll \indic{\{y_1\}}) \,\andthen\,
  (\flipchan \ll \indic{\{y_2\}}) \,\andthen\; \cdots \;\andthen\,
  (\flipchan \ll \indic{\{y_m\}})}.
\end{array} \]

\noindent The $m$-ary conjunction predicate in the latter expression
amounts to $q(x) = x^{n_1}(1-x)^{n_0}$ where $n_{1} = \sum_{i} y_{i}$
is the number of $1$'s among the observation $y_i$ and $n_{0} =
\sum_{i} (1-y_{i})$ is the number of $0$'s, see
Example~\ref{ex:betaflipupdate}. Of course the outcome is
$\betachan(\alpha+n_{1}, \beta+n_{0})$. The question that is relevant
in this setting is: can a random variable $p(y_{1}, \ldots, y_{m})$
with many parameters somehow be simplified, like in $q$ above. This is
where the notion of sufficient statistic arises, see
\textit{e.g.}~\cite{Koopman36,Bishop06}.

\begin{definition}
\label{def:sufstat}
Let $p\colon X\times O^{m} \rightarrow \R$ be a random variable, with
$1+m$ inputs. A \emph{sufficient statistic} for $p$ is a triple of
functions
\[ \xymatrix{
O^{m}\ar[r]^-{s} & \R
\qquad\quad
O^{m}\ar[r]^-{t} & Z
\qquad\quad
X\times Z\ar[r]^-{q} & \R
} \]

\noindent so that $p$ can be written as:
\begin{equation}
\label{eqn:sufstat}
\begin{array}{rcl}
p(x, y_{1}, \ldots, y_{m})
& = &
s(y_{1}, \ldots, y_{m}) \cdot q\big(x, t(y_{1}, \ldots, y_{m})\big).
\end{array}
\end{equation}
\end{definition}

In the above $\betachan$ example we would like to simplify the big
conjunction random variable:
\[ \begin{array}{rcl}
p(x, y_{1}, \ldots, y_{m})
& = &
\Big((\flipchan \ll \indic{\{y_1\}}) \,\andthen\; \cdots \;\andthen\,
  (\flipchan \ll \indic{\{y_m\}})\Big)(x).
\end{array} \]

\noindent We can take $Z = \NNO\times\NNO$ with $t(y_{1}, \ldots,
y_{m}) = (n_{1}, n_{0})$, where $n_1$ and $n_{0}$ are the number of
$1$'s and $0$'s in the $y_i$. Then $q(x, n, n') = x^{n}(1-x)^{n'}$.
The function $s$ is trivial and sends everything to $1$.

A sufficient statistic thus summarises, esp.\ via the function $t$,
the essential aspects of a list of observations, in order to simplify
the update. In the coin example, these essential aspects are the
numbers of $1$'s and $0$'s (that is, of heads and tails).  In these
situations the conjunction predicate --- like $p$ above --- is usally
called a \emph{likelihood}.

The big advantage of writing a random variable $p$ in the form
of~\eqref{eqn:sufstat} is that updating with $p$ can be
simplified. Let $\omega$ be a distribution on $X$, either discrete or
continuous. Then, writing $\vec{y} = (y_{1}, \ldots, y_{m})$ we get:
\[ \begin{array}{rcccl}
\omega|_{p(-,\vec{y})}
& = &
\omega|_{s(\vec{y})\cdot q(-,t(\vec{y}))}
& = &
\omega|_{q(-,t(\vec{y}))}.
\end{array} \]

\noindent The factor $s(\vec{y})$ drops out because it works
like a scalar, see~\eqref{eqn:discconditioningscal}.

We conclude this section with a standard example of a sufficient
statistic (see \textit{e.g.}~\cite{Bishop06}), for a conjunction
expression arising from multiple updates.

\begin{example}
\label{ex:normnormstat}
Recall the $\normchan - \normchan$ conjugate priorship from
Example~\ref{ex:normnorm}. The first channel there has the form
$\normchan = \int u$, for $u(\mu,\sigma,x) =
\nicefrac{1}{\sqrt{2\pi}\sigma}\cdot
e^{-\nicefrac{(x-\mu)^{2}}{2\sigma^2}}$. The second channel is
$\normchan(-,\nu) = \int v$, for a fixed `noise' factor $\nu$, where
$v(x,y) = \nicefrac{1}{\sqrt{2\pi}\nu}\cdot
e^{-\nicefrac{(y-x)^{2}}{2\nu^2}}$. Let's assume that we have
observations $y_{1}, \ldots, y_{m} \in \R_{> 0}$ which we like to use
to iteratively update the prior distribution $\normchan(\mu,\sigma)$.
Following Proposition~\ref{prop:contpriorpointupdate} we
can describe these updates as:
\[ \begin{array}{rcl}
\normchan(\mu,\sigma)\big|_{v(-,y_1)} \;\cdots\; \big|_{v(-,y_m)}
& = &
\normchan(\mu,\sigma)\big|_{v(-,y_1) \,\andthen \;\cdots\; \andthen\,v(-,y_m)}.
\end{array} \]

\noindent Thus we are interested in finding a sufficient statistics
for the predicate:
\[ \begin{array}{rcl}
p(x,y_{1}, \ldots, y_{m})
& \coloneqq &
\Big(v(-,y_1) \,\andthen \;\cdots\; \andthen\,v(-,y_m)\Big)(x)
\\
& = &
v(x,y_{1}) \cdot \;\cdots\; \cdot v(x,y_{m})
\\
& = &
\frac{1}{\sqrt{2\pi}\nu}\cdot e^{-\frac{(y_{1}-x)^{2}}{2\nu^{2}}} \;\cdots\;
   \frac{1}{\sqrt{2\pi}\nu}\cdot e^{-\frac{(y_{m}-x)^{2}}{2\nu^{2}}}
\\
& = &
\frac{1}{(\sqrt{2\pi}\nu)^{m}}\cdot e^{-\frac{\sum_{i}(y_{i}-x)^{2}}{2\nu^2}}
\\
& = &
\frac{1}{(\sqrt{2\pi}\nu)^{m}}\cdot e^{-\frac{\sum_{i}y_{i}^{2}}{2\nu^2}}
   \cdot e^{\frac{2(\sum_{i}y_{i})x-mx^{2}}{2\nu^2}}
\\
& = &
s(y_{1}, \ldots, y_{m}) \cdot q\big(x, t(y_{1}, \ldots, y_{m})\big),
\end{array} \]

\noindent for functions $s,t,q$ given by:
\[ \begin{array}{rclcrclcrcl}
s(y_{1}, \ldots, y_{m}) 
& = &
\frac{1}{(\sqrt{2\pi}\nu)^{m}}\cdot e^{-\frac{\sum_{i}y_{i}^{2}}{2\nu^2}}
& \quad &
t(y_{1}, \ldots, y_{m})
& = &
\sum_{i}x_{i}
& \quad &
q(x, z)
& = &
e^{\frac{2zx-mx^{2}}{2\nu^2}}.
\end{array} \]

\end{example}

\section{Conclusions}\label{sec:conclusions}

This paper contains a novel view on conjugate priors, using the
concept of channel in a systematic manner. It has introduced a precise
definition for conjugate priorship, using a pair of composable
channels $P\rightarrow X\rightarrow O$ and a parameter translation
function $P\times O \rightarrow P$, satisfying a non-trivial equation,
see Definition~\ref{def:conjugateprior}. It has been shown that this
equation holds for several standard conjugate prior examples. There
are many more examples, that have not been checked here. One can be
confident that the same equation holds for those unchecked examples
too, since it has been shown here that conjugate priors amount to
Bayesian inversions. This inversion property is the essential
characteristic for conjugate priors. It has been re-formulated in
logical terms, so that the closure property of a class of priors under
updating is highlighted.

\appendix

\section{Calculation laws for Giry-Kleisli maps with 
   pdf's}\label{sec:calculation}

We assume that for a probability distribution (state)
$\omega\in\Giry(X)$ and a measurable function $f\colon X \rightarrow
\R_{\geq 0}$ the integral $\int f \intd \omega \in [0,\infty]$ can be
defined as a limit of integrals over simple functions that approximate
$f$. We shall follow the description of~\cite{Jacobs13a}, to which we
refer for details\footnote{In~\cite{Jacobs13a} integration $\int
  f\intd \omega$ is defined only for $[0,1]$-valued functions $f$, but
  that does not matter for the relevant equations, except that
  integrals may not exist for $\R_{\geq 0}$-valued functions (or have
  value $\infty$). These integrals are determined by their valued
  $\int \indic{M}\intd \omega = \omega(M)$ on indicator functions
  $\indic{M}$ for measurable subsets, via continuous and linear
  extensions, see also~\cite{JacobsW15a}.}.  This integration
satisfies the Fubini property, which can be formulated, for states
$\omega\in\Giry(X)$, $\rho\in\Giry(Y)$ and measurable function
$h\colon X\times Y \rightarrow \R_{\geq 0}$, as:
\begin{equation}
\label{eqn:productpdfintegration}
\begin{array}{rcl}
\displaystyle\int h \intd (\omega\otimes\rho)
& = &
\displaystyle\int \int h \intd \omega \intd\rho.
\end{array}
\end{equation}

\noindent The product state $\omega\otimes\rho \in \Giry(X\times Y)$
is defined by $(\omega\otimes\rho)(M\times N) =
\omega(M)\cdot\rho(N)$.

\begin{Auxproof}
It suffices to consider the special cases where $g = \indic{M}$ and
$h = \indic{M\times N}$, in:
\[ \begin{array}{rcl}
\displaystyle \int \indic{M} \intd \omega
& = &
\omega(\indic{M})
\\
& = &
\displaystyle \int_{M} f(x)\intd x
\\
& = &
\displaystyle\int f(x) \cdot \indic{M}(x) \intd x
\\
& = &
\displaystyle\int f(x) \cdot g(x) \intd x
\\
\displaystyle \int \indic{M\times N} \intd (\omega\otimes\rho)
& = &
(\omega\otimes\rho)(M\times N)
\\
& = &
\omega(M) \cdot \rho(N)
\\
& = &
\big(\displaystyle\int_{M} f(x) \intd x\big)\cdot 
   \big(\displaystyle \int \indic{N} \intd \rho\big)
\\
& = &
\displaystyle \int \big(\displaystyle\int_{M} f(x) \intd x\big)\cdot \indic{N} 
   \intd \rho
\\
& = &
\displaystyle \int \int \indic{M}(x)\cdot f(x)\cdot \indic{N}  \intd x
   \intd \rho
\\
& = &
\displaystyle \int \int \indic{M\times N}(x,-)\cdot f(x) \intd x \intd \rho
\\
& = &
\displaystyle \int \int h(x,-)\cdot f(x) \intd x \intd \rho.
\end{array} \]
\end{Auxproof}

\subsection{States via pdf's}\label{subsec:stateviapdf}

For a subset $X\subseteq \R$, a measurable function $f\colon X
\rightarrow \R_{\geq 0}$ is called a probability density function
(pdf) for a state $\omega\in\Giry(X)$ if $\omega(M) = \int_{M} f(x)
\intd x$ for each measurable subset $M\subseteq X$. In that case we
simply write $\omega = \int f(x) \intd x$, or even $\omega = \int
f$. If $\omega$ is given by such a pdf $f$, integration with state
$\omega$ can be described as:
\begin{equation}
\label{eqn:pdfintegration}
\begin{array}{rcl}
\displaystyle\int g \intd \omega
& = &
\displaystyle\int f(x)\cdot g(x) \intd x.
\end{array}
\end{equation}

\subsection{Channels via pdf's}\label{subsec:channelviapdf}

Let channel $c\colon X \rightarrow \Giry(Y)$ be given as $c = \int u$
by pdf $u\colon X\times Y \rightarrow \R_{\geq 0}$ as $c(x)(N) =
\int_{N} u(x,y) \intd y$, for each $x\in X$ and measurable $N\subseteq
Y$, like in~\eqref{eqn:channelfrompdf}. If $\omega = \int f$ is a
state on $X$, then state transformation $c \gg \omega \in \Giry(Y)$
is given by:
\begin{equation}
\label{eqn:pdfstatetransformation}
\hspace*{-0.8em}\begin{array}{rcl}
(c \gg \omega)(N)
\hspace*{\arraycolsep}\smash{\stackrel{\eqref{eqn:statetransformation}}{=}}\hspace*{\arraycolsep}
\displaystyle \int c(-)(N) \intd \omega
& \smash{\stackrel{\eqref{eqn:pdfintegration}}{=}} &
\displaystyle \int f(x) \cdot c(x)(N) \intd x
\\
& = &
\displaystyle \int_{N} \int f(x) \cdot u(x,y) \intd x \intd y.
\end{array}
\end{equation}

\noindent Hence the pdf of the transformed state $c \gg \omega$ is $y
\mapsto \int f(x) \cdot u(x,y) \intd x$.

Given a channel $d \colon Y \rightarrow \Giry(Z)$, say with $d = \int
v$, then sequential channel composition $d \after c$ is given, for
$x\in X$ and $K\subseteq Z$, by:
\begin{equation}
\label{eqn:pdfseqcomposition}
\begin{array}{rcl}
(d \after c)(x)(K)
\hspace*{\arraycolsep}\smash{\stackrel{\eqref{eqn:continuouscomposition}}{=}}\hspace*{\arraycolsep}
\displaystyle \int d(-)(K) \intd c(x)
& \smash{\stackrel{\eqref{eqn:pdfintegration}}{=}} &
\displaystyle \int u(x,y) \cdot d(y)(K) \intd y
\\
& = &
\displaystyle \int_{K} \int u(x,y) \cdot v(y,z) \intd y \intd z
\end{array}
\end{equation}

\noindent We see that the pdf of the channel $d \after c$ is $(x,z)
\mapsto \int u(x,y) \cdot v(y,z) \intd y$.

For a channel $e = \int w \colon A \rightarrow \Giry(B)$ we get
a parallel composition channel $c\otimes e \colon X\times A
\rightarrow \Giry(Y\times B)$ given by:
\begin{equation}
\label{eqn:pdfparcomposition}
\begin{array}{rcl}
(c \otimes e)(x,a)(M\times N)
& = &
c(x)(M) \otimes e(a)(N)
\\
& = &
\displaystyle\big(\int_{M} u(x,y)\intd y\big)\cdot
   \big(\int_{N} w(a,b) \intd b\big)
\\[0.8em]
& = &
\displaystyle \int_{M\times N} u(x,y)\cdot w(a,b) \intd (y,b).
\end{array}
\end{equation}

\noindent Hence the pdf of the channel $c\otimes d$ is $(x,a,y,b)
\mapsto u(x,y)\cdot w(a,b)$.

\subsection{Graph channels and pdf's}\label{subsec:graphandpdf}

For a channel $c\colon X \rightarrow \Giry(Y)$ we can form `graph'
channels $\tuple{\idmap,c} = (\idmap\otimes c) \after \copier \colon X
\rightarrow \Giry(X\times Y)$ and $\tuple{c,\idmap} = (c\otimes\idmap)
\after \copier \colon X \rightarrow \Giry(Y\times X)$. For $x\in X$ we
have:
\begin{equation}
\label{eqn:graphequation}
\begin{array}{rclcrcl}
\tuple{\idmap,c}(x)
& = &
\eta(x)\otimes c(x)
& \qquad\mbox{and}\qquad &
\tuple{c,\idmap}(x)
& = &
c(x) \otimes \eta(x).
\end{array}
\end{equation}

\begin{Auxproof}
\[ \begin{array}{rcl}
\tuple{\idmap, c}(x)(M\times N)
& = &
\big(\st_{2} \after (\idmap\times c) \after \Delta\big)(x)(M\times N)
\\
& = &
\big(\dst \after (\eta\times c) \after \Delta\big)(x)(M\times N)
\\
& = &
\dst\big((\eta\times c)(x,x)\big)(M\times N)
\\
& = &
\dst\big(\eta(x), c(x)\big)(M\times N)
\\
& = &
\eta(x)(M) \cdot c(x)(N)
\\
& = &
\big(\eta(x) \otimes c(x)\big)(M\times N).
\end{array} \]
\end{Auxproof}

\noindent If $c = \int u$ and $\omega = \int f$ is a state on $X$, then:
\begin{equation}
\label{eqn:pdfgraphstatetransformation}
\begin{array}{rcl}
(\tuple{\idmap,c} \gg \omega)(M\times N)
& \smash{\stackrel{\eqref{eqn:pdfintegration}}{=}} &
\displaystyle \int f(x) \cdot \tuple{\idmap,c}(x)(M\times N) \intd x
\\
& \smash{\stackrel{\eqref{eqn:graphequation}}{=}} &
\displaystyle \int f(x) \cdot \eta(x)(M) \cdot c(x)(N) \intd x
\\
& = &
\displaystyle \int_{N} \int_{M} f(x) \cdot u(x,y) \intd x \intd y.
\end{array}
\end{equation}

\noindent We also consider the situation where $d\colon X\times Y
\rightarrow \Giry(Z)$ is of the form $d = \int v$, with composition:
\begin{equation}
\label{eqn:pdfgraphcomposition}
\begin{array}{rcl}
\big(d \after \tuple{\idmap,c}\big)(x)(K)
& \smash{\stackrel{\eqref{eqn:graphequation}}{=}} &
\displaystyle \int d(-)(K) \intd (\eta(x)\otimes c(x))
\\
& \smash{\stackrel{\eqref{eqn:productpdfintegration}}{=}} &
\displaystyle \int d(-)(K) \intd \eta(x) \intd c(x)
\\
& = &
\displaystyle \int d(x,-)(K) \intd c(x)
\\
& \smash{\stackrel{\eqref{eqn:pdfintegration}}{=}} &
\displaystyle \int u(x,y) \cdot d(x,y)(K) \intd y
\\
& = &
\displaystyle \int_{K} \int u(x,y) \cdot v(x,y,z) \intd y \intd z.
\end{array}
\end{equation}

\noindent Hence the pdf of the channel $d \after \tuple{\idmap,c}$
is $(x,z) \mapsto \int u(x,y) \cdot v(x,y,z) \intd y$.

%\bibliography{/home/bart/svn/bart/Tex/bib}

\begin{thebibliography}{}

\bibitem[Abramsky and Coecke, 2009]{AbramskyC09}
Abramsky, S. and Coecke, B. (2009).
\newblock A categorical semantics of quantum protocols.
\newblock In Engesser, K., Gabbay, D.~M., and Lehmann, D., editors, {\em
  Handbook of Quantum Logic and Quantum Structures: Quantum Logic}, pages
  261--323. North-Holland, Elsevier, Computer Science Press.

\bibitem[Ackerman et~al., 2011]{AckermanFR11}
Ackerman, N., Freer, C., and Roy, D. (2011).
\newblock Noncomputable conditional distributions.
\newblock In {\em Logic in Computer Science}. IEEE, Computer Science Press.

\bibitem[Alpaydin, 2010]{Alpaydin10}
Alpaydin, E. (2010).
\newblock {\em Introduction to Machine Learning}.
\newblock {MIT} Press, Cambridge, MA, $2^{\textrm{nd}}$ edition.

\bibitem[Bernardo and Smith, 2000]{BernardoS00}
Bernardo, J. and Smith, A. (2000).
\newblock {\em Bayesian Theory}.
\newblock John Wiley \& Sons.

\bibitem[Bishop, 2006]{Bishop06}
Bishop, C. (2006).
\newblock {\em Pattern Recognition and Machine Learning}.
\newblock Information Science and Statistics. Springer.

\bibitem[Blackwell, 1951]{Blackwell51}
Blackwell, D. (1951).
\newblock Comparison of experiments.
\newblock In {\em Proc. Sec. Berkeley Symp. on Math. Statistics and
  Probability}, pages 93--102. Springer/British Computer Society.

\bibitem[Cho and Jacobs, 2017]{ChoJ17b}
Cho, K. and Jacobs, B. (2017).
\newblock The {EfProb} library for probabilistic calculations.
\newblock In Bonchi, F. and K{\"o}nig, B., editors, {\em Conference on Algebra
  and Coalgebra in Computer Science (CALCO 2017)}, volume~72 of {\em LIPIcs}.
  Schloss Dagstuhl.

\bibitem[Cho et~al., 2015]{ChoJWW15b}
Cho, K., Jacobs, B., Westerbaan, A., and Westerbaan, B. (2015).
\newblock An introduction to effectus theory.
\newblock see \url{arxiv.org/abs/1512.05813}.

\bibitem[Clerc et~al., 2017]{ClercDDG17}
Clerc, F., Dahlqvist, F., Danos, V., and Garnier, I. (2017).
\newblock Pointless learning.
\newblock In Esparza, J. and Murawski, A., editors, {\em Foundations of
  Software Science and Computation Structures}, number 10203 in Lect. Notes
  Comp. Sci., pages 355--369. Springer, Berlin.

\bibitem[Culbertson and Sturtz, 2014]{CulbertsonS14}
Culbertson, J. and Sturtz, K. (2014).
\newblock A categorical foundation for bayesian probability.
\newblock {\em Appl. Categorical Struct.}, 22(4):647--662.

\bibitem[Diaconis and Ylvisaker, 1979]{DiaconisY79}
Diaconis, P. and Ylvisaker, D. (1979).
\newblock Conjugate priors for exponential families.
\newblock {\em Annals of Statistics}, 7(2):269--281.

\bibitem[Faden, 1985]{Faden85}
Faden, A. (1985).
\newblock The existence of regular conditional probabilities: Necessary and
  sufficient conditions.
\newblock {\em The Annals of Probability}, 13(1):288--298.

\bibitem[Giry, 1982]{Giry82}
Giry, M. (1982).
\newblock A categorical approach to probability theory.
\newblock In Banaschewski, B., editor, {\em Categorical Aspects of Topology and
  Analysis}, number 915 in Lect. Notes Math., pages 68--85. Springer, Berlin.

\bibitem[Jacobs, 2013]{Jacobs13a}
Jacobs, B. (2013).
\newblock Measurable spaces and their effect logic.
\newblock In {\em Logic in Computer Science}. IEEE, Computer Science Press.

\bibitem[Jacobs, 2015]{Jacobs15d}
Jacobs, B. (2015).
\newblock New directions in categorical logic, for classical, probabilistic and
  quantum logic.
\newblock {\em Logical Methods in Comp. Sci.}, 11(3).
\newblock See \url{https://lmcs.episciences.org/1600}.

\bibitem[Jacobs, 2017]{Jacobs17a}
Jacobs, B. (2017).
\newblock From probability monads to commutative effectuses.
\newblock {\em Journ. of Logical and Algebraic Methods in Programming},
  94:200--237.

\bibitem[Jacobs and Westerbaan, 2015]{JacobsW15a}
Jacobs, B. and Westerbaan, A. (2015).
\newblock An effect-theoretic account of {Lebesgue} integration.
\newblock In Ghica, D., editor, {\em Math. Found. of Programming Semantics},
  number 319 in Elect. Notes in Theor. Comp. Sci., pages 239--253. Elsevier,
  Amsterdam.

\bibitem[Jacobs and Zanasi, 2016]{JacobsZ16}
Jacobs, B. and Zanasi, F. (2016).
\newblock A predicate/state transformer semantics for {Bayesian} learning.
\newblock In Birkedal, L., editor, {\em Math. Found. of Programming Semantics},
  number 325 in Elect. Notes in Theor. Comp. Sci., pages 185--200. Elsevier,
  Amsterdam.

\bibitem[Koopman, 1936]{Koopman36}
Koopman, B. (1936).
\newblock On distributions admitting a sufficient statistic.
\newblock {\em Trans. Amer. Math. Soc.}, 39:399--409.

\bibitem[Panangaden, 2009]{Panangaden09}
Panangaden, P. (2009).
\newblock {\em Labelled {Markov} Processes}.
\newblock Imperial College Press, London.

\bibitem[Russell and Norvig, 2003]{RussellN03}
Russell, S. and Norvig, P. (2003).
\newblock {\em Artificial Intelligence. A Modern Approach}.
\newblock Prentice Hall, Englewood Cliffs, NJ.

\bibitem[Selinger, 2007]{Selinger07}
Selinger, P. (2007).
\newblock Dagger compact closed categories and completely positive maps
  (extended abstract).
\newblock In Selinger, P., editor, {\em Proceedings of the 3rd International
  Workshop on Quantum Programming Languages (QPL 2005)}, number 170 in Elect.
  Notes in Theor. Comp. Sci., pages 139--163. Elsevier, Amsterdam.
\newblock DOI \url{http://dx.doi.org/10.1016/j.entcs.2006.12.018}.

\bibitem[Selinger, 2011]{Selinger11}
Selinger, P. (2011).
\newblock A survey of graphical languages for monoidal categories.
\newblock In Coecke, B., editor, {\em New Structures in Physics}, number 813 in
  Lect. Notes Physics, pages 289--355. Springer, Berlin.

\bibitem[Stoyanov, 2014]{Stoyanov14}
Stoyanov, J. (2014).
\newblock {\em Counterexamples in Probability}.
\newblock Wiley, $2^{\textrm{nd}}$ rev. edition.

\end{thebibliography}

\end{document}